\documentclass[apjl]{emulateapj}

\usepackage{amsmath}
\usepackage{amssymb}
\usepackage{amsfonts}
\usepackage{amstext}
\usepackage{graphicx}
\usepackage{fancybox}
\usepackage[usenames,dvipsnames]{color}
\usepackage{pstricks}
\usepackage{bm}
\usepackage{soul, color}
\usepackage{latexsym}
\usepackage{pifont}
\usepackage{shorttoc}
\usepackage{subfigure}
\usepackage{textcomp} 
\usepackage{float}
\usepackage{enumitem}

\usepackage{bm}
\usepackage{lipsum}
\usepackage{threeparttable}
\usepackage{hyperref}
\hypersetup{urlcolor=cyan}

\def\aap{A\&A}

\def\apjs{ApJS}
\def\apjl{ApJL}

\newcommand{\lya}{Ly$\alpha$\ }

\newcommand{\angs}{\, {\rm \AA}}

\newcommand{\no}[1]{}

\newcommand{\myemail}{l.m.ribas@astro.uio.no}
\def\lsim{~\rlap{$<$}{\lower 1.0ex\hbox{$\sim$}}}

\def\gsim{~\rlap{$>$}{\lower 1.0ex\hbox{$\sim$}}}

\shorttitle{Revealing the Warm and Hot Halo Baryons via Thomson Scattering of Quasar Light}
\shortauthors{Llu\'is Mas-Ribas \& Joseph F. Hennawi}

\begin{document}

\title{Revealing the Warm and Hot Halo Baryons via Thomson Scattering of Quasar Light}

\author{Llu\'is Mas-Ribas\altaffilmark{1}} 
\author{Joseph F. Hennawi\altaffilmark{2,3}} 
\altaffiltext{1}{Institute of Theoretical Astrophysics, University of Oslo,
Postboks 1029, 0315 Oslo, Norway.\\
 \url{\myemail}}
\altaffiltext{2}{Department of Physics, University of California, Santa 
Barbara, CA 93106, USA.}
\altaffiltext{3}{Max-Planck-Institut f{\"u}r Astronomie, K{\"o}nigstuhl 17, 
D-69117 Heidelberg, Germany.}


\begin{abstract}  

The baryonic content and physical properties of the warm and hot
($10^5\lesssim T\lesssim 10^7$ K) phases of the circumgalactic medium
(CGM) are poorly constrained, owing to the lack of observables
probing the requisite range of temperature, spatial scale, halo mass, and redshift.
The radiation from a luminous quasar produces a spatially extended emission halo
resulting from Thomson scattering off of free electrons in the CGM, which       
can be used to measure the electron density profile, and therefore, the amount
of warm and hot baryonic matter present.  We predict the resulting
surface brightness profiles and show that they are easily detectable in a three hour 
integration with the James Webb Space Telescope (JWST), 
out to $\sim 100$ physical kpc from the centers of individual hyper-luminous quasars.
This electron scattering surface brightness is redshift independent,
and the signal-to-noise ratio depends only very weakly on redshift,
in principle allowing measurements of the warm and hot CGM into the Epoch of
Reionization at $z\sim 6.5$. We consider a litany of potential
contaminants, and find that for fainter quasars at $z\lesssim1$,
extended stellar halos might be of comparable surface brightness. 
At $z>2$, JWST mid-IR observations start to probe rest-frame
optical/UV wavelengths implying that scattering by dust grains in the
CGM becomes significant, although multi-color observations should be
able to distinguish these scenarios given that Thomson scattering
is achromatic.

\end{abstract}

\section{Introduction}

   The circumgalactic medium (CGM) is the region extending up to a few hundreds of kiloparsecs
around galaxies, and where the  interactions between galaxies and the intergalactic medium (IGM) 
take place. The gas that fuels star formation is accreted from the IGM onto the galaxy, and  
the material processed in the interstellar medium (ISM) can be expelled toward its outskirts 
in galactic winds. As the CGM constitutes the primary flow of baryonic matter in and out galaxies, 
its study is crucial for understanding galaxy formation and evolution. 

    The rest-frame ultraviolet (UV) absorption features that the CGM gas produces in the 
spectra of background sources have been used to probe this medium 
for about half a century \citep[see, e.g.,][for a review]{Tumlinson2017}, but advances in observational 
instrumentation and methodology are now also enabling detailed studies in emission
\citep{Steidel2011,HennawiProchaska2013,Cantalupo2014,Hennawi2015,Arrigonibattaia2015}.
Recent observations indicate that most star-forming galaxies at high
redshift show extended emission in their CGM, usually in the form of
Ly$\alpha$, from several tens up to $\sim 80-100$ pkpc \citep[e.g.,][]{Matsuda2012,
Momose2014,Wisotzki2016,Leclercq2017,Xue2017}, and often also in H$\alpha$, up to a 
few tens of kpc from the central stellar regions \citep{Hayes2013,Matthee2016,Sobral2016}.
This diffuse emission is a new window into the structure of the CGM, and can provide unique 
information about faint star formation in the halo of galaxies \citep{Masribas2017}, as well as the 
escape of ionizing photons from galaxies into the IGM up to the redshifts of cosmic reionization 
\citep{Masribas2017b}. For the case of bright quasars (or active galactic nuclei; AGNs), their 
ionizing radiation can illuminate dense gas in their surroundings, resulting in even larger 
Ly$\alpha$ nebulosities that extend up to a few hundreds of kpc \citep[e.g.,][]{Prescott2009,
Yang2009,HennawiProchaska2013,Arrigonibattaia2014,Cantalupo2014,Martin2014,
Roche2014,Hennawi2015,Arrigonibattaia2016,Borisova2016}. In extreme cases, this 
phenomenon can trace the densest environments, providing a signpost for the
most massive (proto-) galaxies and clusters at $z \sim 2-3$ \citep{Hennawi2015,Martin2015,
Cai2017,Arrigonibattaia2018}. 

The aforementioned observables provide valuable information about the medium within the 
temperature range $\sim10^4$ and a few times $10^5$ K (which we will hereafter refer to as the
{\it cool phase}),
but a complete picture of the CGM also includes gas at temperatures $10^5\lesssim T 
\lesssim 10^7$ K. Hydrodynamical simulations show that the majority of the baryons
interior to the virial radius of massive halos ($M_{\rm h} \gtrsim
10^{12}\,{\rm M_{\odot}}$) are shock-heated to the virial temperature
$T_{\rm vir} \gtrsim 10^6\,{\rm K}$ ({\it hot phase}; \citealt{Birnboim2003}), but the 
detection of these baryons, and especially of those in the range $10^5\lesssim 
T \lesssim 10^6$ K ({\it warm phase}), is difficult. Constraining the amount of warm and 
hot baryonic matter confined within  halos is crucial to ({\it i}) gain insight into the relevance 
of feedback processes; if feedback effects are small, we expect most of the virialized 
gas to remain in the halo, whereas, if feedback is important, this gas will be 
expelled into the IGM and mixed with the warm-hot intergalactic medium 
\citep[WHIM;][]{Cen2006,Roncarelli2012,Vandevoort2016}. ({\it ii}) Quantifying the  
amount of warm and hot gas in galactic halos is important for shedding light on the
so called 'missing baryons problem',  which is that only a small fraction of the total 
baryon budget  (inferred from the Cosmic Microwave Background, CMB, and Big Bang nucleosynthesis) 
have been detected thus far
\citep{Persic1992,Fukugita1998,Fukugita2004,Tripp2004,Prochaska2008,Shull2012}.
Recent studies 
of cosmic filaments by \cite{Tanimura2017} and \cite{Degraaff2017} suggest that $\gtrsim 30 \%$ of 
the missing baryons is in the WHIM, consistent with the 
$50 \%$ inferred from simulations by \citealt{Hojjati2015}, but the halo component is much more 
uncertain \citep[e.g.,][]{Anderson2010,Mcgaugh2010}. To assess this baryonic content, current 
studies typically make use of observations of the Sunyaev-Zel'dovich effect and X-rays,  but 
these observables are not sensitive to the entire range of spatial scales, temperatures, and/or 
redshifts covered by the warm and hot phases of the CGM \citep[see the review 
by][]{Bregman2007}.  

     X-ray observations only enable studies of the hot component, at $T\gtrsim 10^6$ K, and the signal 
is heavily weighted toward small-scale regions because it depends on the square of the density 
\citep{Gupta2012}, and the metallicity \citep{Bogdan2017,Li2017}, both of which
rise steeply
toward the halo center. At redshifts $z\gtrsim 0.5$, the expected X-ray emission is too faint to 
detect galactic halos and, even at redshifts as low as $z\lesssim0.1$, the detection is 
challenging. Diffuse emission from the halos of individual massive ($M_{\rm h} \gsim10^{13}
\,{\rm M_{\odot}}$) spiral and early-type elliptical galaxies at $z\lesssim0.1$ has been 
detected in a few tens of objects, but only out to a few tens of kpc from their 
centers \citep{Anderson2011,Dai2012x,Humphrey2012, Bogdan2013,Anderson2016,
Goulding2016,Bogdan2017,Li2017}. In most of these cases it is challenging to separate 
the signal from the background beyond $\sim10-20$ kpc, and   
for the compact inner emission, it is unclear whether it  arises from 
the hot gas or is produced by faint discrete sources of stellar nature \citep[e.g., X-ray 
binaries and cataclysmic processes;][]{Bogdan2012,Dai2012x}. Additional sensitivity has    
been obtained by stacking large samples of objects, enabling detections of halo emission beyond 
a few hundreds of kpc from the centers of massive galaxies ($M_{\rm h}\gsim10^{12.7}\,{\rm 
M_{\odot}}\,$\footnote{We have converted the halo masses to the virial halo mass using the 
mass-conversion relations by \cite{Hu2003} when they are defined in another nomenclature.}) 
typically residing in the centers of galaxy clusters \citep[e.g.,][]{Anderson2013,
Anderson2015}. To summarize, X-ray studies are limited to low redshifts, large masses and hot 
gas, and for individual objects, to small scales around the galactic centers.

    Large distances from the centers of individual massive galaxy groups and clusters can be reached by 
analysing the thermal Sunyaev-Zel'dovich effect \citep[tSZ;][]{Sunyaev1970,sunyaev1972} that energetic  
free electrons in the CGM have on CMB photons. Contrary to the case of X-rays, the tSZ signal is linear in 
the electron density, and hence less weighted toward the center. Furthermore, the tSZ effect is in principle 
redshift independent, although in practice size evolution and limited spatial resolution render objects 
at redshifts $z\gsim1.5$ currently undetectable
(see the reviews by \citealt{Carlstrom2002} and \citealt{Kitayama2014}). 
Individual objects (clusters) with halo masses in the range $M_{\rm h}\sim2-5\times10^{14}\,{\rm M_{\odot}}$ 
can be detected up to $z\sim0.3$, and up to $z\sim1.5$  for larger masses \citep{Bleem2015}, but stacking 
(or cross-correlating) thousands of objects enhances the sensitivity and, therefore, enables detections of 
less massive halos and higher redshifts \citep{Scannapieco2008}. The \cite{Planck2013}, \cite{Greco2015}, 
and \cite{Ruan2015} cross-correlated tSZ maps with $z\sim 0.03$ locally bright galaxies (LBGs) from 
the Sloan Digital Sky Survey DR7 \citep[SDSS;][]{Abazajian2009}, allowing them to assess the signal 
in halos of masses down to $M_{\rm h}\sim4.4\times10^{12}\,{\rm M_{\odot}}$. \cite{Spacek2016,
Spacek2017} used tSZ stacks to study the environment around $M_{\rm 
  h}\sim7\times10^{13}\,{\rm M_{\odot}}$ elliptical galaxies in the redshift range $z \in [0.1 - 1.5]$,
and 
\cite{Hand2011} and \cite{Chatterjee2010} stacked the signal around 
$M_{\rm h}\sim 10^{14}\,{\rm M_{\odot}}$ luminous 
red galaxies (LRGs) within $0.16 \lesssim z \lesssim 0.47$. Finally, \cite{Gralla2014} studied the 
tSZ signal around radio galaxies, with average halo masses $M_{\rm h}\sim 10^{13}\,{\rm M_{\odot}
}$ and median redshift $z\sim1$. Overall, these studies detected average warm and hot phases broadly 
consistent with the amount expected from the theory for virialized galactic halos.

Stacking the tSZ signal around large samples of bright quasars allows one
to characterize the properties of high-redshift halos, as well as the
impact of quasar feedback on their CGM.  \cite{Chatterjee2010} and
\cite{Ruan2015} performed such analyses with SDSS quasars covering the
redshift range $0.08 \lesssim z \lesssim 2.82$, enabling them to probe
halo masses  $M_{\rm h} > 10^{12.5}\,{\rm M_{\odot}}$.
These studies found a total thermal energy in the halos
exceeding the values expected from gravitational heating by up to $\sim1$
order of magnitude, which indicates a strong contribution from
feedback. However, \cite{Cen2015} argued that the observed thermal
energies can be fully explained via gravitational heating alone, given
the large beam sizes of a few arcmin for WMAP and Planck, and
the uncertainties in the dust temperature in the calibration of the tSZ
maps. More recently, \cite{Crichton2015} performed tSZ analyses
similar to those by Chatterjee et al. and Ruan et al. with a
smaller-beam experiment (ACT, $\sim 1$ arcmin; \citealt{Swetz2011})
and also found a large thermal energy excess, consistent with the 
results by \cite{Dutta2017}, although the importance of feedback still 
remains under debate because of the difficulties in
analysing and interpreting the tSZ signal in all these analyses 
\citep[][see also the recent findings by \citealt{Spacek2017b}]{Lebrun2015,Verdier2016,Hill2017}. In detail,
extracting conclusive information from the observations of the tSZ
effect is difficult because the separation between the actual tSZ
signal and that from other contaminants, i.e., thermal radiation from
dust, requires a precise (not straightforward) modeling of the
emission spectrum at various frequency bands
\citep{Cen2015,Greco2015,Ruan2015}. Furthermore, the signal is
unresolved at distances within the beam of the instrument and only the
integrated effect, characterized by the Compton-$y$ parameter, can be
measured. Finally, there is a degeneracy between the 
electron temperature and density, $y\propto n_{\rm e}(r)\,T_{\rm
  e}(r)$, which could vary radially, and that  
requires additional modeling  and/or assumptions
(e.g., an isothermal medium) to separate these dependencies and obtain
quantitative constraints.

The observation of the diffuse extended emission around a luminous
quasar/AGN that results from nuclear light that has been scattered by
the free electrons (Thomson scattering) in the host CGM is another
potential probe of the baryons residing in galactic halos
\citep{Sunyaev1982,Sholomitskii1990}.  Obvious advantages of this
approach compared to other observables are: (${\it i}$) this effect is
sensitive to the presence of all the baryons in the halo, irrespective
of their temperature, provided that they are ionized; (${\it
  ii}$) In constrast with X-ray studies, the signal is linearly
proportional to the electron density, $n_{\rm e}(r)$, resulting in a
signal that decays more gradually with radius, implying potentially
detectable emission at large distances; (${\it iii}$) quantifying the
implied CGM density profiles is straightforward because there is no
degeneracy with other parameters, in contrast with both X-ray and tSZ
studies.

Because scattered radiation is also polarized
\citep[e.g.,][]{Lee1999}, electron scattering has been invoked to
explain the diffuse polarized continuum emission in the central
regions ($\lesssim 1$ kpc) of nearby AGNs by \cite{Antonucci1985,
  Koyama1989,Miller1990,Antonucci1994,Ogle2003}. On larger scales, i.e., 
a few tens of kpc from the central source, scattering has been suggested
as a potential mechanism to explain the polarization around
radio galaxies at redshifts  $0.5 \lesssim z \lesssim 
2$ by \cite{Dey1996,Tran1998,Cohen1999}, but in these cases 
whether the scattering medium was electrons or dust was
unclear \citep[see also][]{Kishimoto2001,Vernet2001}.  \cite{Geller2000} attempted
to detect extended halos of polarized radio emission around a bright
$z\sim3$ radio galaxy resulting from the Thomson scattering of nuclear
emission, but the limited sensitivity resulted only in weak upper
limits for the halo/IGM baryon content.  However, theoretical predictions
by \cite{Holder2004} showed that modern radio interferometers may be
able to detect this signal around bright radio sources inhabiting the
centers of massive clusters. In view of these results, it seems that
electron scattering has not yet provided competitive constraints on halo
baryons because limited statistics and sensitivity imply that the majority
of detections are limited to small scales, where the particle dominating
the scattering process (dust or electrons) is unclear.

In principle, the nature of the scattering medium can be easily determined given
a spectrum of the scattered radiation.  When quasar radiation is
scattered, the photons inherit a Doppler frequency shift resulting from the velocities of
the scattering particles, implying that the emission lines will
be broadened by thermal velocity dispersion of the scattering medium.
Because this thermal line broadening scales as the scattering particle mass, 
$m^{-1/2}$, the large masses of dust grains will result in negligible broadening.  
For electron scattering, however, the resulting velocity width of scattered
quasar emission lines is of the order $\gsim \sqrt{m_{\rm p}/m_{\rm e}}\,v_{\rm vir} =
10^4\,{\rm km\, s^{-1}}$ in massive galaxies and clusters
\citep{Loeb1998}, exceeding the typical intrinsic quasar line values
of $\sim 5\times 10^3\,{\rm km\, s^{-1}}$ \citep{Peterson1997}, where $m_{\rm p}$ and 
$v_{\rm vir}$ are the proton mass and virial velocity, respectively.
This effect has been explored in the early theoretical papers by
\cite{Gilfanov1987,Fabian1989,Sarazin1993} to discuss the beamed AGN
radiation. Since the broadening is linearly proportional to the
thermal velocity of the gas, \cite{Loeb1998} and \cite{Khedekar2014}
showed that analysis of the line width can also be used to derive the
electron temperature of the gas, which is a
valuable complementary probe, together with the diffuse emission, to
better constrain the properties of the warm and hot gas.

We propose here calculations of the extended radiation of a
hyper-luminous quasar scattered by the 
free electrons in the CGM of the
host galaxy, and demonstrate that this emission is a viable and unique
tool to probe the properties of the warm and hot gas in the halo that is difficult
to detect via other methods. The emission
profiles appear to be potentially detectable with NIRCam onboard JWST,
and spatially resolved out to large radial scales, enabling studies of
individual halos at masses lower than X-ray and tSZ approaches, and at
redshifts up to those of the Cosmic Reionization. Furthermore, we stress that the signal is
independent of temperature and linearly proportional to electron
density, enabling 
constraints on the total baryon content.

In \S~\ref{sec:formalism}, we detail the formalism for the calculation
of the surface brightness profiles of electron and dust scattering,
and in \S~\ref{sec:media} we present our simple model for
the medium around the
quasar-host galaxy. In \S~\ref{sec:contaminants} we explore the
potential contaminants for the scattering signal, and detail our
observational strategy in \S~\ref{sec:obs}.  We present the results in
\S~\ref{sec:results}, and discuss our findings and approach in
\S~\ref{sec:discussion}, before concluding in
\S~\ref{sec:conclusions}.

We assume a flat $\Lambda$CDM cosmology with the parameter values from \cite{Planck2015}.

\section{Formalism}\label{sec:formalism}

   We consider a two phase CGM, which for simplicity we call hot and cool phases, 
the latter containing dust, and calculate the extension of the quasar emission scattered by the 
free electrons and the dust in these media, respectively. Modeling the signal from dust is important 
given that it can potentially contaminate that from electrons. 

   We present in \S~\ref{sec:sb} the formalism for the calculation of the surface brightness profiles that 
result from the scattering processes of dust and electrons. In \S~\ref{sec:scat}, we detail the calculations 
of the scattering redistribution function for electrons (\S~\ref{sec:escat}), and dust (\S~\ref{sec:dscat}). 

\subsection{Scattering Surface Brightness Profile}\label{sec:sb}

   The surface brightness profile at impact parameter $r{_\perp}$ from the central source 
results from integrating the radiation scattered in the host halo along the line-of-sight $s$ 
toward the observer. 
The radial coordinate $r$ is related to $r{_\perp}$ and $s$ as $r^2=r_{\perp}^2 + s^2$, 
so that $r\,{\rm d}r = s\,{\rm d}s$, and the surface brightness can be expressed as  
\citep{Masribas2016}
\begin{align}\label{eq:jint}
{\rm SB}(r_{\perp},\nu_{\rm obs})= & \frac{1}{(1+z)^3}\int j(r,\nu_0){\rm d}s \\ \nonumber 
 = &\frac{2}{(1+z)^3}\int_{r_{\perp}}^{\infty}  j(r,\nu_0) 
	\frac{r\, {\rm d}r}{\sqrt{r^2 - {r_{\perp}}^2}} ~ .
\end{align}
Here, $\nu_{\rm obs}$ is the frequency of the radiation in the observer's frame, which is
related to the  frequency  $\nu_0$ emitted in the rest-frame of the source via $(1+z)\nu_{\rm obs} =\nu_0$, 
where $z$ denotes the redshift of the source. The term $1/(1+z)^3$ results from cosmological dimming of 
the spectral brightness ${\rm SB}(r_{\perp},\nu_{\rm obs})$, and $j(r,\nu_0)$ denotes the volumetric emissivity of 
the scattered radiation at frequency $\nu_0$ and distance $r$, which can be further specified as
\begin{align}
  j(r,\nu_0) = & {n_x(r)\sigma_{\rm x}} \frac{L_{\nu_0}}{4\pi r^2}P_x \\ \nonumber
              = & {\tau_{x,0}} f_{{\rm V},x}(r_{\rm vir})\frac{L_{\nu_0}}{4\pi r_{\rm vir}^3} \left(\frac{r_{\rm vir}}{r}\right)^{\alpha_x + 2} P_x  ~. 
\end{align}
The factor $1/{4\pi r^2}$ above accounts for the geometric dilution of the specific luminosity, $L_{\nu_0}$, of the 
central source at frequency $\nu_0$. The terms $n_x(r)$ and $\sigma_x$ denote the radial volume density 
profile and the scattering cross section, respectively, of the scattering particles, and $x$ takes on `hot' 
(`e'; electrons) or `cool' (`d'; dust) regarding the two CGM phases (scatterers) considered in our model 
(\S~\ref{sec:gal}). The function $P_x$ denotes the integral of the scattering redistribution function 
$R(\nu,\Omega)_x$, over both solid angle $\Omega$ (between our line-of-sight and the original photon 
emission direction), and (original) emission frequency $\nu$, i.e. $P_x= \int \int R(\nu,\Omega)_x \,{\rm d}\nu 
\, {\rm d}\Omega$ (\S~\ref{sec:scat}). We express the emissivity in terms of the scattering optical depth, 
$\tau_{x,0}$, considering $\tau_{x,0}=n_{x,0} \sigma_x r_{\rm vir}$\,\footnote{In detail, the quantity $\tau_{x,0}$ 
represents the differential optical depth at the virial radius since $\tau_x(R) = \int_0^R {\rm d}\tau_x(r)=\int_0^R 
n_x(r)\sigma_x{\rm d}r$. We use this term as a parameterization, and avoid the differential notation for 
simplicity.}, and parameterizing the density as $n_x(r)=n_{x,0}f_{{\rm V},x}(r_{\rm vir})\left(\frac{r_{\rm 
vir}}{r}\right)^{\alpha_x}$, where $n_{x,0}$  and $f_{{\rm V},x}(r_{\rm vir})$ are the density and volume filling 
factor values, respectively, at the virial radius $r_{\rm vir}$, and $\alpha_x$ is the power-law index of the profile. 
Combining the previous two equations, the resulting surface brightness finally equals
\begin{align}\label{eq:sb}
  {\rm SB}(r{_\perp},\nu_{\rm obs}) & =  \frac{1}{(1+z)^3} \tau_{x,0} f_{{\rm V},x}(r_{\rm vir}) \frac{L_{\nu_0}}{2\pi 
  r_{\rm vir}^2} \\  \nonumber 
  &  \times  \int_{r{_\perp}}^{\infty} P_x
  \left(\frac{r_{\rm vir}}{r}\right)^{\alpha_x+1} \frac{{\rm
      d}r}{\sqrt{r^2- r{_\perp}^2}} ~ .
\end{align}

   We emphasize that  the linear dependence of the electron-scattered surface brightness profile 
on the electron density is encapsulated in the optical depth, i.e., $\tau_{\rm hot} \propto n_{\rm e}$, 
and describe in more detail the parameters of these equations below.

\subsection{Scattering Redistribution Function}\label{sec:scat}

For most scattering processes, the absorbed radiation is not
re-emitted isotropically but rather in preferred directions that
depend on the nature of the scattering medium. The probability that
the scattered photons are directed along a given direction is
represented by the redistribution (or phase) function
$R(\nu',\bf{n}';\nu,\bf{n})$, which denotes the probability of
scattering a photon from a frequency $\nu'$ to $\nu$ and from a
direction $\bf{n}'$ to $\bf{n}$ \citep{Dirac1925}. For our purposes,
the vector $\bf{n}$ will denote the direction along the line-of-sight
toward an observer on Earth, whereas the vector $\bf{n}'$ indicates
the direction of the radiation emitted by the quasar.
We describe the phase
function for electron scattering in \S~\ref{sec:escat} and for dust in
\S~\ref{sec:dscat}. In all cases, we assume that the photons undergo
only a single scattering event, which is a valid approximation 
since the CGM is optically thin to scattering by dust
and electrons as shown in \S~\ref{sec:gal}.

\subsubsection{Hot CGM - Electron scattering}\label{sec:escat}

   For the hot phase, we consider that the electron temperature is similar to the virial temperature 
${ T_{\rm vir} \sim 10^6}$ K, implying that the free electrons are in the non-relativistic regime 
($k_{\rm B}T_{\rm e}\ll{m_{e} c^2}$, where $k_{\rm B}$ denotes the Boltzmann constant, 
$m_e$ the electron mass, and $c$ is the speed of light), and that the photon energies 
are below X-ray energies of $0.511$ MeV, implying the low-energy 
scattering limit (${ h_{\rm P}}\nu \ll {m_e c^2}$, where $h_{\rm P}$ is Planck's constant). 
In this regime,  and assuming that the free electrons follow a Maxwellian velocity distribution,
the phase function is the classical Thomson redistribution function \citep{Rybicki1979,Loeb1998}\footnote{While our equation is adequate 
for our purposes, we refer the interested reader to \cite{Sazonov2000}, for a detailed and comprehensive 
work on the redistribution function for high energy photons and semi-relativistic cases.} of the form
\begin{align}\label{eq:rede}
R(\nu',\bf{n}';\nu,{\bf{n}})_{\rm hot} &= \frac{\rm 3}{\rm 4}(1+\mu^2)\,\frac{1}{[2\pi\beta_T^2(1-\mu)\nu^2]^{1/2}}~ \nonumber \\
						 &\times\, {\rm exp}\left[-\frac{(\nu-\nu')^2}{2\beta_T^2(1-\mu)\nu^2}\right]~,
\end{align}
where $\mu=\bf{n}'\cdot\bf{n}=\cos {\rm \Theta}$, with ${\rm \Theta}$ the angle between the vectors $\bf{n}'$ 
and $\bf{n}$, 
and $\beta_T^2\equiv2k_{\rm B}T_{\rm e}/m_{\rm e}c^2$, 
such that $\beta_T=v_{\rm th}/c$ is the electron thermal velocity
in units of $c$. For the present purposes, 
we will be considering observations through broad-band filters, and so the frequency redistribution will
just broaden the underlying quasar spectrum by a small amount compared to the filter widths we consider. 
Therefore, we ignore the frequency dependence of the redistribution function and 
integrate between the limits of the frequency range covered by the 
filter\footnote{We have ignored the filter curve in this integral since this would mostly only 
change the normalization constant.} used in our default observational settings (\S~\ref{sec:obs}), and normalize 
the phase function such that  $\int \int C\,R(\nu',{\bf{n}}';\nu,{\bf{n}})_{\rm hot}\,{\rm d}\nu\,{\rm 
d}\Omega=1$, where $C$ is a normalization constant.

 \begin{figure*}\center 
\includegraphics[width=0.80\textwidth]{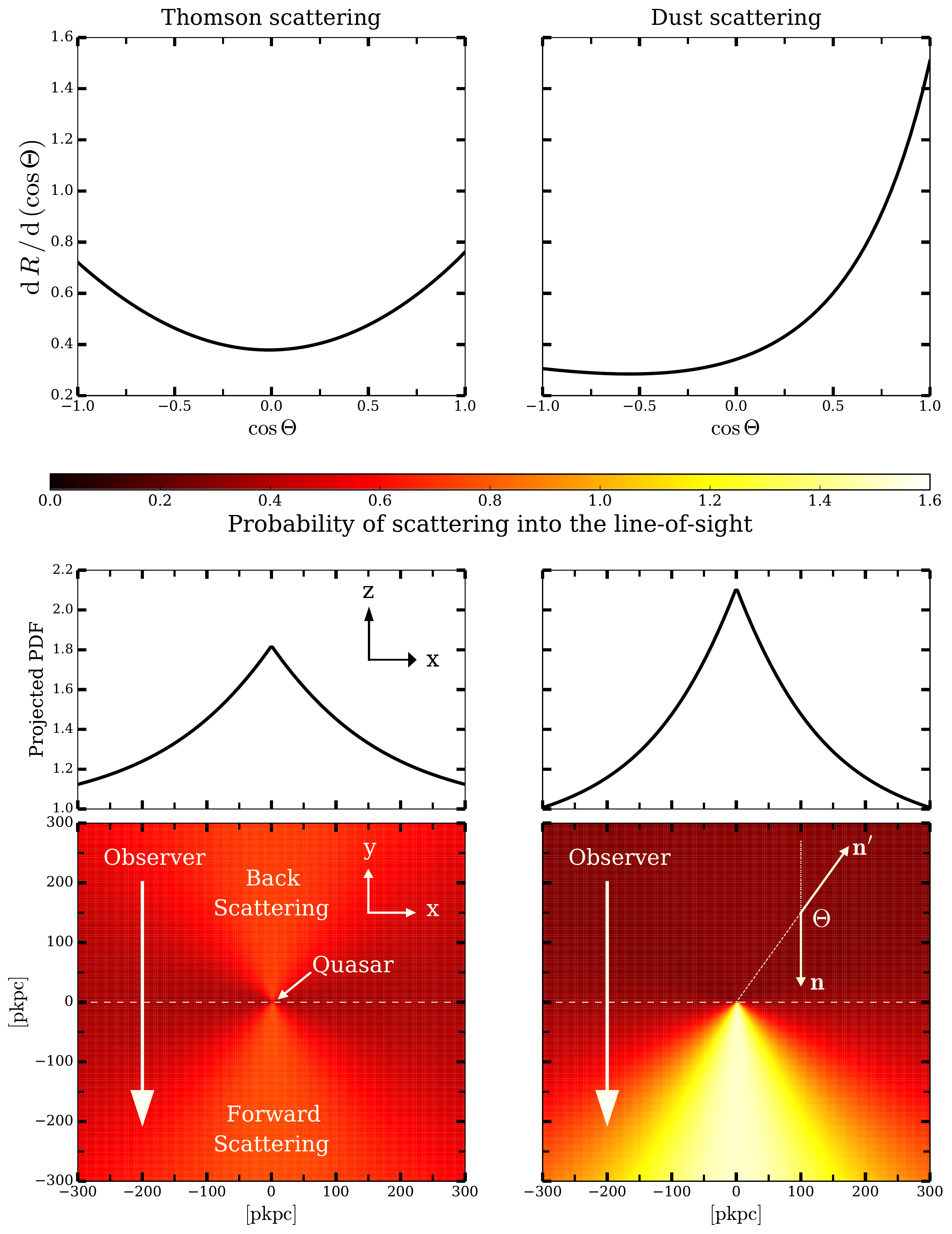}
\caption{Nature and impact of the scattering redistribution functions, of non-relativistic free electrons ({\it left 
column}), and dust ({\it right column}). The {\it top panels} represent the redistribution functions with 
$\cos\Theta$, where $\Theta$ is the angle between the incoming and outcoming photon. The right panel 
shows the preference for forward scattering for the case of dust. The {\it lower panels} 
display the probability distribution functions of scattering photons toward the observer at every position of a  
plane $\bf{n}\times \bf{n}'$ containing the quasar in the center. The {\it middle panels} show the previous 
distributions projected 
along the observer's line-of-sight, indicating that the highest projected probability of scattering photons 
toward the observer are found at small impact parameters from the quasar. In the bottom right panel, we 
illustrate an example of the angle $\Theta$ between the line-of-sight and the radial vector for a given position of 
the plane.} 
\label{fig:redist}
\end{figure*}

   The {\it top left panel} in Figure \ref{fig:redist} displays the electron redistribution function with $\mu=\cos 
\Theta$, showing that the forward- and back-scattering scenarios are favored ($\mu=1$ and $\mu=-1$, 
respectively), while the lowest probability is for $\mu=0$ ($\Theta=90^\circ$ between the incoming and 
outcoming photon directions). The forward scattering is slightly more favored than the backward one.
The {\it bottom left panel} represents the probability of scattering a photon into the line-of-sight toward the 
observer for every position in a plane intersecting the quasar host, defined by $\bf{n}\times \bf{n}'$, 
the vector normal to the plane, where $\bf{n}$ defines the direction toward the observer, and $\bf{n}'$ is
the direction of ray of radiation emitted by the quasar.  
The {\it middle left panel} shows the projection along the
line-of-sight ($\bf{n}$) of the 2D distribution, which will
be the relevant quantity for the surface brightness calculation. Since
forward and back scattering are preferred, this leads to the
projection of the redistribution function peaking at small impact
parameters, because at these distances most of the contribution comes
from photons emitted in the direction parallel (or antiparallel) to
the observer. At large impact parameters, the line-of-sight integral
has a much greater contribution from photons emitted at intermediate
angles, which have lower probability of scattering toward the
observer.

\subsubsection{Cool CGM - Dust scattering}\label{sec:dscat}

   For the cool phase, we consider the scattering by the dust particles embedded in this medium. The 
Henyey-Greenstein function \citep{Henyey1941} is usually used to describe the anisotropic scattering 
phase function of a mixture of dust grains, but \cite{Draine2003} proposed an improved function to better 
match the real one at wavelengths $\lambda>1\,\mu{\rm m}$. We adopt the dust redistribution function from
\citet{Draine2003}
\begin{equation}\label{eq:redd}
R(\mu)_{\rm cool} = \frac{1}{4\pi}\frac{1-g^2}{1+\alpha(1+2g^2)/3}\frac{1+
\alpha\mu^2}{(1+g^2-2g\mu)^{3/2}}~,
\end{equation}
with the parameters $g=0.26$ and $\alpha=0.62$. These parameter values result from considering a Milky 
Way dust model and radiation at $\lambda=1.2\,\mu{\rm m}$ (lower panel of Figure 6 in 
\citealt{Draine2003}), consistent with our approach for dust (next section) and our proposed observational setup
(\S~\ref{sec:obs}). Considering these parameters and the term $1/4\pi$ in Eq.~\ref{eq:redd}, this function 
does not require further normalization.

   The dust redistribution function is plotted in the {\it top right panel} of Figure \ref{fig:redist}, which shows 
the strong preference of dust for scattering the radiation in the same direction as the incoming 
photons (forward scattering), maximizing the probabilities at angles $|\Theta| \lesssim 30^\circ$. The favored 
forward-scattering scenario is clearly observed in the 2D PDF plot in the {\it bottom right panel} of Figure 
\ref{fig:redist}, appearing as a bright triangular area in the lower half of the plot, while the upper 
part, representing back scattering, is almost homogeneously dark. This dust property 
results in a more sharply peaked projected probability profile compared to that of electrons 
({\it middle right panel}).

\section{Parameterization of the host CGM}\label{sec:media}
 
    In this section, we describe how we model the physical properties of the CGM of the host galaxy    
(\S~\ref{sec:gal}), which we use to compute 2D maps of scattered quasar emission 
(\S~\ref{sec:2dmap}). 

\subsection{Host Galaxy CGM}\label{sec:gal}

   We consider a two-phase halo, consisting of a hot CGM phase composed of hot plasma that has 
been shock heated to the halo virial temperature and is collisionaly ionized  (\S~\ref{sec:hotp}), and 
a cool (dusty) CGM component, characterized in the works of the Quasars Probing Quasars (QPQ) series 
by \cite{Hennawi2006}, \cite{Hennawi2007}, \cite{PH2009}, \cite{Lau2016}, 
and by \lya emission constraints from the work by \citealt{Arrigonibattaia2016}
(\S~\ref{sec:coldp}).

\subsubsection{Hot CGM phase}\label{sec:hotp}

   We model the distribution of hot gas as 
\begin{equation}\label{eq:hotgas}
n_{\rm H,hot}(r) = n_{\rm H,hot,0}\,f_{\rm V, hot}(r_{\rm vir})\left(\frac{r}{r_{\rm vir}}\right)^{-\alpha_{\rm h}}~,
\end{equation}
where we assume a volume filling factor $f_{\rm V, hot}(r_{\rm vir})=1$, the term $\alpha_{\rm h}=5/2$  is derived 
from the hydrodynamical simulations by \cite{Nelson2016}, and $n_{\rm H,hot,0}$ is the 
volume density of hydrogen in the hot phase at the virial radius, obtained as follows.    
In Nelson et al., the dark matter 
halos with mass $M_{\rm h}\sim10^{12}\,M_{\odot}$ at $z=2$ have a virial radius $r_{\rm vir}\sim 
100$ pkpc and a gas density  $n_{\rm H,0}\sim10^{-3.75}\,{\rm cm^{-3}}$, which 
we use to calculate the hot gas mass enclosed within their virial radius as $M_{\rm vir,hot}=4\pi\, 
n_{\rm H,0}f_{\rm V, hot}(r_{\rm vir})\,r_{\rm vir}^3/(3-\alpha_{\rm h})/X$, with $X=0.76$ denoting the cosmic hydrogen 
abundance. The ratio between the hot and total baryonic mass is then  
obtained using the cosmic baryon fraction, $f_{\rm b}=\Omega_{\rm b}/\Omega_{\rm m}=
0.174$, resulting in $M_{\rm vir,hot}/(f_{\rm b}M_{\rm h}) = 0.83$, which we fix for our further calculations. 
In practice, variations in redshift, halo mass, galaxy type, etc., may change the value of the ratio, 
but we expect the halo of massive quasar hosts to be dominated by hot gas in any case, as 
indicated by simulations \citep[e.g.,][]{Birnboim2003,Nelson2016}.
We have checked that variations of this 
value by $\lesssim 25 \%$ do not greatly alter our conclusions. Finally, we obtain 
$n_{\rm H,hot,0}$ by solving the above equations using now the fixed ratio, a halo mass 
$M_{\rm h}=10^{12.5}\,M_{\odot}$, characteristic of dark matter halos hosting quasars 
\citep[][see also \citealt{Conroy2013}]{White2012}, and the corresponding virial radius 
for this halo mass at the redshift of interest. 

This hot phase contains free electrons that will scatter the quasar radiation. We parameterize 
the Thomson scattering optical depth by these free electrons with the quantity $\tau_{\rm hot,0}=n_{\rm 
e,0}\sigma_{\rm e}r_{\rm vir}$, where $\sigma_{\rm e} = \sigma_T \equiv 6.65\times10^{-25}\,{\rm cm^{2}}$ 
is the Thomson scattering cross-section, and $n_{\rm e,0}=(1+Y/2X)n_{\rm H,hot,0}$ is the electron volume 
density at $r_{\rm vir}$, with $Y=0.24$ denoting the cosmic helium abundance.
In general in our calculations, the photons reaching the CGM have traversed an electron-scattering 
optical depth with a value $\sim10^{-2}$, consistent with the typical value of the intra-cluster medium. 
This optical depth represents an optically thin medium to electron scattering that ensures the validity of the 
single-scattering approximation in our calculations. We neglect the scattering driven by dust in this phase because we expect a dust-to-gas mass ratio far 
below 1\% \citep[][and references therein]{Laursen2010}.

\subsubsection{Cool CGM phase}\label{sec:coldp}

    We characterize the cool CGM phase with a population of $T\sim10^4$ K spherical gas clouds of 
uniform density, $n_{\rm H,cool,0}$, distributed in the halo according to a radial volume filling factor 
of the form \citep{HennawiProchaska2013}
\begin{equation}\label{eq:vfil}
f_{\rm V,cool}(r) = f_{\rm V,cool}(r_{\rm vir})\left(\frac{r}{r_{\rm vir}}\right)^{-\alpha_{\rm c}}~,
\end{equation}
where $f_{\rm V,cool}(r_{\rm vir})$ is the volume filling factor at the virial radius, and $\alpha_c$ is the 
power-law index of the density scaling relation in this medium. Similarly, the density distribution of cool gas 
with radial distance from the center can be expressed as 
\begin{equation}\label{eq:coolgas}
n_{\rm H,cool}(r) = n_{\rm H,cool,0}\,f_{\rm V,cool}(r_{\rm vir})\left(\frac{r}{r_{\rm vir}}\right)^{-\alpha_{\rm c}}~.
\end{equation}

Given these assumptions, the average column density of cool gas at impact parameter 
$r{_\perp}$ is obtained as 
\begin{align}\label{eq:colden}
\langle N_{\rm H,cool}&(r_{\perp})\rangle = \int_{r_{\perp}} n_{\rm H,cool}(r){\rm d}s~ \nonumber \\
		      &= 2N_{\rm H,cool,0}\sqrt{\left(\frac{r_{\rm max}}{r_{\rm vir}}\right)^2  - \left(\frac{r_{{\perp}}}{r_{\rm vir}}\right)^2} \left(\frac{r_{\rm \perp}}{r_{\rm vir}}\right)^{\rm -\alpha_h}~ \nonumber \\
		      &\times  {_2F_1} \left[\frac{1}{2},\frac{\rm \alpha_h}{2}, \frac{3}{2},1 - \left(\frac{r_{\rm max}}{r_{\perp}}\right)^2\right] ~, 
\end{align}
where $r_{\rm max}=2\,r_{\rm vir}$ is the maximum radius out to which the profile in Eq.~\ref{eq:coolgas} 
extends, $N_{\rm H,cool,0}$ denotes the cool gas column density at the virial radius, and the term $_{2}F_1$ 
is the Gaussian hypergeometric function that accounts for the integral along 
the line-of-sight at given impact parameter. A similar expression to Eq.~\ref{eq:colden} 
holds for the mean column density of the hot phase, $\langle N_{\rm H,hot}(r_{\perp})\rangle$, provided 
we replace $\alpha_{\rm c}$ by $\alpha_{\rm h}$ and $N_{\rm H,cool,0}$ by $N_{\rm H,hot,0}$.

We set $\alpha_{\rm c} = 0$, owing to the weak radial dependence of $N_{\rm H}$ on impact 
parameter out to $\approx 200\,{\rm kpc}$ obtained by \cite{Lau2016}, from the photoionization 
modeling of a sample of  
$z\sim 2-3$ background quasar sightlines passing through the CGM. Since we previously 
obtained that $83\%$ of the baryonic mass is in the hot gas, we determine the degenerate 
product $n_{\rm H,cool,0}f_{\rm V,cool}(r_{\rm vir})$ from the expression $M_{\rm vir,cool}=4\pi n_{\rm H,cool,0}
f_{\rm V,cool}(r_{\rm vir})(\,r_{\rm vir}^3/(3-\alpha_{\rm c})/X$ by assuming that the rest 17$\%$ 
of the mass is in the cool phase. This calculation results in a value of the column density at the virial radius of 
$N_{\rm H,cool,0}\sim 10^{20}\,{\rm cm^{-2}}$, broadly consistent with the results 
by \cite{Lau2016} and \cite{PH2009}.

 \begin{figure*}\center 
\includegraphics[width=0.97\textwidth]{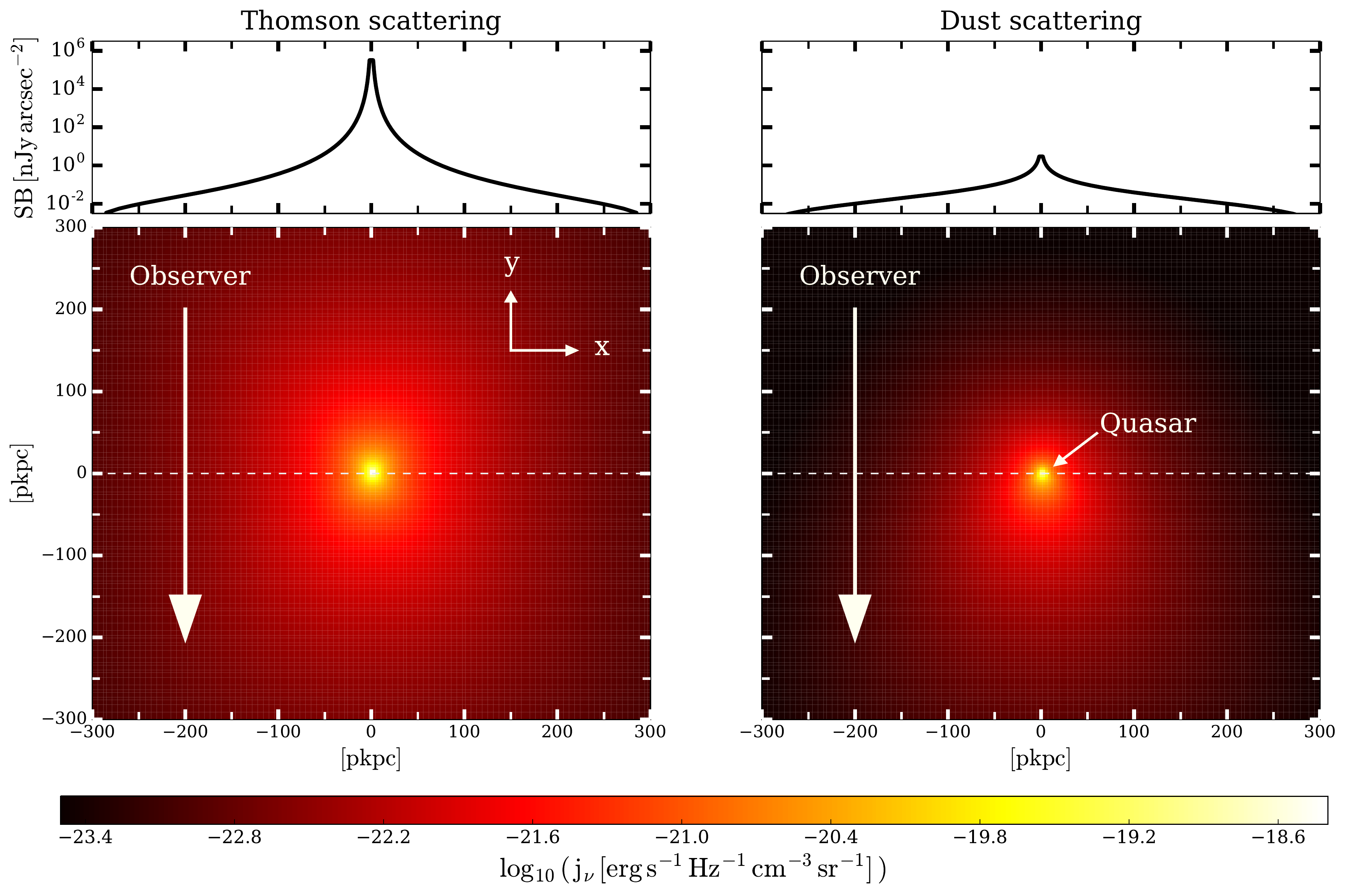}
\caption{{\it Bottom panels}: two-dimensional maps of quasar radiation scattered into the line-of-sight  toward 
the observer at every point 
of the $\bf{n}\times \bf{n}'$ plane. The {\it left panel} illustrates the the elongated distribution of scattered 
emission driven 
by the Thomson scattering redistribution function. The {\it right panel} shows the preference by the dust 
redistribution function for forward scattering, which favors the emission from the scattering sites placed 
at small impact parameters from the central quasar. {\it Top panels:} surface brightness profiles 
of the lower panels integrated along the line-of-sight for each case.}
\label{fig:2dmap}
\end{figure*}

The scattered emission from the cool phase is driven by the presence of dust in the CGM, which 
has been inferred at large radial distances, from several tens of kpc up to several Mpc from the centers of 
$z\sim0.3$ SDSS galaxies by \citealt{Menard2010} \citep[see also][]{Peek2015}. We quantify 
the dust-scattering optical depth assuming that the dust-to-gas ratio scales as the CGM metallicity 
compared to the solar value, $Z_{\odot}$, with a ratio $Z/Z_{\odot}=1/10$ \citep{Prochaska2013b}, 
and a reddening parameter $R_{\rm V}=3.1$, the value typically considered for the ISM of the Milky Way 
with a column density $\sim 2\times10^{21}\,{\rm cm^{-2}}$. Using these parameters and our 
cool gas column density results in a reddening ${\rm E(B-V)}=\frac{Z}{Z_{\odot}}\,\frac{N_{\rm 
H,cool,0}}{2\times10^{21}\,{\rm cm^{-2}}}\sim 10^{-2}$, in agreement with the weak evidence for 
reddening of SDSS quasars reported by \citealt{Krawczyk2015} \citep[see also][]{Richards2003}. 
We adopt the Milky Way extinction curve by \cite{Cardelli1989}, which gives the attenuation curve 
A$_\lambda$, because it 
extends into the near infra-red (NIR; considered for our observations), although the curve for the 
Small Magellanic Cloud (SMC) might be a more accurate choice given the low metallicity of the CGM 
\citep{Hutchings1982,Laursen2010,Peek2015}. However, we have tested that extrapolating the SMC law by 
\cite{Gordon2003} to the NIR does not alter our results. We set the dust albedo (the probability that a photon is 
scattered instead of destroyed by dust) to $a_{\rm d}=0.7$, consistent with the values by \cite{Li2001} and 
\cite{Draine2003}. Finally, the dust optical depth parameter equals 
$\tau_{\rm cool,0,\lambda}=a_{\rm d}\,{\rm A_{\lambda}}\,\frac{R_{\rm V}}{3.1} \,\frac{Z}{Z_{\odot}}\,
\frac{N_{\rm H,cool,0}}{2\times10^{21}\,{\rm cm^{-2}}}$, which, considering the rest-frame NIR  
wavelengths of our fiducial calculations results in dust optical depths of the order $10^{-5}$, about 
three orders of magnitude smaller than that of electrons.

\subsection{Maps of Scattered Emission}\label{sec:2dmap}

Once the parameterization of the CGM and the redistribution functions
are established, we can calculate the emissivity of radiation
scattered into the line-of-sight toward the observer at every position
of the halo.  The {\it bottom panels } in Figure \ref{fig:2dmap}
display this spatial distribution in the plane defined by
$\bf{n}\times \bf{n}'$ for electrons ({\it left panel}), and dust
({\it right panel}).  The colors indicate the value of the emissivity
at every point of the plane, with brighter colors denoting higher
values. The {\it left panel} shows a distribution elongated along the
line-of-sight, driven by the preference of Thomson scattering for
forward and back scattering, with the signal decreasing with impact
parameter from the center. The {\it right panel} indicates that most
of the signal arises from radiation scattered between the quasar and
the observer, at small angles from the line-of-sight ($\lesssim
60^\circ$), resulting from the forward-scattering redistribution
function pattern of dust. Most of the radiation emitted at larger
angles does not contribute to the observed signal, except for
scatterings occurring close to the center ($r\lesssim 50 - 70$
pkpc). The {\it upper panels} show the surface brightness profiles
resulting from integrating the scattered emissivity of the lower 2D
maps along the line-of-sight toward the observer (i.e.,
Eq.~\ref{eq:jint}). For this calculation, we have used the parameters
of the central source that we detail in \S~\ref{sec:obs}. Briefly, we
have considered the rest-frame radiation at $1.8\,\mu$m from a
hyper-luminous quasar at $z=1$, with an apparent magnitude in the
$i$-band of $15.5$ mag.

\section{Other Sources of Diffuse Halo Emission}\label{sec:contaminants}

    The electron-scattering surface brightness profile could be potentially confused with the
signal from dust,  but also with other sources of diffuse emission from the halo of the host galaxy.
We thus also estimate the  potential contamination from the nebular emission resulting from the interaction between 
the quasar radiation and the cool CGM gas in \S~\ref{sec:nebular}, and that by star 
formation in the host galaxy in \S~\ref{sec:halosf}.
   
\subsection{Extended Nebular Emission}\label{sec:nebular}

   We estimate the level of nebular emission in the halo, which arises from the 
interaction between the radiation of the central quasar and the cool gas. We use the photoionization code 
{\sc Cloudy} v10.01, last described by \cite{Ferland2017}, and follow the method by 
\citealt{Arrigonibattaia2015} (their sections 4.3 and 4.4), for which we briefly describe the main 
steps below. We use the quasar spectral energy distribution (SED) described in \S~\ref{sec:obs} 
for our adopted central source as input in {\sc Cloudy} to parameterize the radiation field 
illuminating the CGM, and the values for the cool CGM phase detailed in the 
previous sections to characterize the gas. Given the small size of the cool clouds 
compared to their distance from the central source (more than $\sim$50 times larger), we 
assume a plane-parallel geometry for the computations. The output from {\sc Cloudy} is the emissivity $j_{\nu}$
arising from the cool clouds at various distances from the central source, from which we 
compute the surface brightness at various impact parameters using Eq.~\ref{eq:jint}. 
Despite its simplicity, this calculation is enough to ensure that the nebular emission is 
not a contaminant of the electron-scattered signal in the first $150-200$ pkpc. 
We show the nebular surface brightness profile, together with those of dust and electrons, 
in \S~\ref{sec:profiles}. 

\subsection{Stellar Emission}\label{sec:halosf}

 Stellar emission in the quasar host halo could be a significant contaminant 
of the electron-scattering signal. We estimate the impact of this effect accounting for 
the signal from possible extended stellar halos in \S~\ref{sec:star}, and satellite 
galaxies clustered around the central source in \S~\ref{sec:sat}.

\subsubsection{Extended Stellar Halos}\label{sec:star}

     Observations of massive $\log (M_{\rm h}/{\rm M_{\odot}})\gsim 12.5 - 13$ early-type elliptical 
galaxies at redshifts $z\lesssim 1$ show extended stellar halos that can be individually detected 
out to several tens of kiloparsecs \citep[e.g.,][]{Schombert2015,Buitrago2017,Huang2017,Oh2017}, 
and up to a few hundreds of kpc when stacking a large number of them \citep[e.g.,][]{Kormendy2009,
Tal2011,Souza2014}, with radial profiles significantly flatter than those of other types of galaxies, 
e.g., spirals, that typically have more compact stellar components \citep[e.g.,][]{Courteau2011}. 
The origin of these extended stellar halos is an unsolved problem, but it has been suggested that
they arise from rapid growth of the progenitor galaxies 
at redshifts $z\sim2-4$ \citep[e.g.][]{Dekel2009}, followed by the quenching of star-formation driven 
by stellar and/or AGN feedback \citep[see][for a review]{Harrison2017}, and a final 
period (at $z\lesssim1$) of non-dissipative merger events with other galaxies \citep{Conroy2007,
Purcell2007,Szomoru2012,Patel2013}. 

The diffuse emission from these extended stellar halos could be comparable to
the electron scattering signal if the quasar host resembles a massive elliptical galaxy,
which is often the case for luminous quasars at low-redshift \citep[$z\lesssim 1$; 
e.g.,][]{Guyon2006,Hyvonen2007,Hyvonen2007b,Veilleux2009}.
However, observational characterization of these extended stellar
surface brightness profiles is challenging because it often requires
assumptions about the intrinsic ellipticity of the galaxy, complex
subtraction of contaminant neighboring and/or foreground sources, as
well as modeling of the the point-spread function (PSF) of the
observations \citep{Abraham2017, Knapen2017}. Because observations
of extended stellar halos are available at low redshift where they are also expected to be larger,
we estimate the intensity of this signal in our discussion of the electron scattering
emission from the quasar 3C 273 at $z\sim0.16$ in  \S~\ref{sec:3c273}.

\subsubsection{Satellite Galaxies}\label{sec:sat}

The presence of faint galaxies around the quasar can be another
potential contaminant to the scattering emission. If these galaxies
are individually detectable, they can be masked and removed from the
images but, otherwise, we cannot separate their contribution from the
diffuse halo emission. For the case of undetectable galaxies along the
line-of-sight but far from the host halo, we expect their distribution
to be uncorrelated with the quasar, and therefore their emission is
effectively part of the sky background which will be subtracted from
images of the quasar.  However, faint (undetectable) galaxies
clustered around the central quasar produce a signal which cannot be
masked, and which will not subtract out.  Recent studies have
indicated that the clustering of galaxies around quasar hosts is
significant
\citep[e.g.,][]{Coil2007,Trainor2012,Wang2015,Decarli2017,Garcia2017}.
Thus, the collective, azimuthally integrated emission from these clustered
satellite sources can produce diffuse light in the host halo, as we
discussed in \cite{Masribas2016} and \cite{Masribas2017}, which could
masquerade as the scattering signal we are interested in. Below we estimate
the size of this effect. 

The first step in this calculation is to obtain the luminosity
(magnitude) above which satellite galaxies clustered around the
central quasar can be individually detected, and therefore removed,
given our observational setup. We discuss the observational strategy
in detail in \S~\ref{sec:obs}, but we briefly present the essential
details here.  We consider observations using the NIRCam
instrument onboard the James Webb Space Telescope
\citep[JWST;][]{Gardner2006}, with the broad-band filter F356W, which
covers the spectral region where the sky background is minimum. The
default exposure time is set to 7200s, and we assume the quasar host
halo to be at redshift $z=1$. This redshift implies observing NIR
emission from the sources (centered at $\sim 1.8\,\mu$m in the source
rest-frame), a region of the galaxy spectrum dominated by stellar
continuum light. We consider that a source is detectable when its
signal is, at least, five times the value of the noise $\sigma$, with
$\sigma = \sqrt{N_{\rm sky}+{\rm RN}^2}$. Here, $N_{\rm sky}$ and
${\rm RN=2}$ are the photon counts for the sky brightness and
instrumental readout noise, respectively, detailed in the next
section. In the calculation of the sky photon count we have considered
an effective area for the satellite galaxies $S_{\rm eff}=\pi r_{\rm
  eff}^2$, where $r_{\rm eff}=0.5$ arcsec is the galaxy radius. We
plug these numbers into Eq.~\ref{eq:Nobs} and obtain the minimum flux
of a detectable galaxy, i.e., our detection threshold, resulting in
$m_{\rm AB}\sim28.3$ AB apparent magnitudes, broadly consistent with
the sensitivity estimates in the NIRCam documentation
\footnote{\url{https://jwst-docs.stsci.edu/display/JTI/NIRCam+Imaging+Sensitivity} 
(Figure 1, scaled to our $5\sigma$ value)}. 

As shown below, our calculations result in an average number of $\sim 0.25$ undetectable satellite 
galaxies in the halo, which imply highly stochastic surface brightness values that depart strongly 
from the deterministic average profile. We assess the impact of the satellites using a 
Monte Carlo approach to better capture this effect, instead of using the analytical approach that we adopted 
in \cite{Masribas2017}, where we analytically computed the mean (field) emissivity of the faint galaxy 
population and boosted its value close to the central source using the correlation function. In detail, 
we now perform $10^6$ realizations where, in each of them, we populate the host 
halo with satellite galaxies by randomly sampling the luminosity and correlation functions that describe the    
satellite population. When a satellite galaxy is above our detection threshold, we consider that it  
can be detected and masked out, and we do not add it to the halo. Therefore, each realization is 
a possible scenario illustrating the impact of the undetectable satellite sources on the scattering surface 
brightness profiles. We characterize the 
parameters of the luminosity function for satellite sources by using the fitting formula in \citealt{Stefanon2013} 
(their section 4.2) at redshift $z=1$, which is constrained at the rest-frame H-band ($\sim 1.6\,\mu$m), 
consistent with the rest-frame wavelength of our observations. The absolute magnitude of our detection 
threshold then corresponds to $M_{\rm H,thr}=-15.9$ mag assuming a negligible $K$-correction 
\citep{Pogi1997,Mannucci2001}, 
and the parameters of the luminosity function, $\phi_{\rm H}(M)$, are 
$M_{\rm H}^*=-23.88\,{\rm mag}$, $\phi_{\rm H}^*=1.1\times10^{-3}{\rm \,mag^{-1}\,Mpc^{-3}}$, and power-law 
index $\alpha_{\rm H}=-1.15$. The observed magnitudes in \citealt{Stefanon2013} cover the range 
$-M_{\rm H}\sim 18 - 26$ mag, but we extend these limits for our calculations. We set the upper 
limit of the integral over luminosity function to $M_{\rm H,max}=-27$ mag for numerical purposes, although the exact
value is irrelevant given that 
these satellites are above the detection threshold and will not be considered. For the lower limit, we integrate
down to $M_{\rm H,min}=-12$ mag accounting for possible undetected galaxies not captured in the luminosity 
functions by \cite{Stefanon2013}. We have tested that variations around this limit do not alter our results 
because the faint-end slope of the luminosity function is significantly flat. 

In order to model the clustering of satellites around the central quasar, 
we use the power-law cross-correlation function between galaxies and quasars at $z=1$ reported by 
\cite{Coil2007}, $\xi^{\rm GQ}(r)=(r/r^{\rm GQ}_0)^{\gamma^{\rm GQ}}$, with scale-length $r^{\rm GQ}_0=
3.3\, h^{-1}\,{\rm cMpc}$ and power-law index $\gamma^{\rm GQ}=-1.55$. We divide our range of 
magnitudes in 35 bins, resulting in variations ${\rm d}M_{\rm H}=0.43$ mag, and the radial distances 
between $r_{\rm min}=20$ kpc and $r_{\rm max}=300$ kpc in 19 evenly distributed logarithmic bins, 
resulting in ${\rm d}\log (r/{\rm kpc})=0.06$. The lower limit for the radial distance is set considering that 
the possible effect of the quasar host galaxies are not captured by our simple method. 
The chosen number of bins allow us to precisely sample the distributions 
while not slowing down the computations; we have tested that the results are 
insensitive to the exact number of bins. We finally populate the halos following the steps 
described below: 
\begin{enumerate}[leftmargin=0pt,itemindent=20pt]
\item We calculate the exact number of galaxies at every radial bin $i$ with the expression
	\begin{equation}
	n_{i} = \int_{M_{\rm H,min}}^{M_{\rm H,max}} \phi_{\rm H}(M)\,{\rm d}M \int_{r_i}^{r_{i+1}}4\pi 
		r^2 \,[1+\xi^{\rm GQ}(r)]\,{\rm d}r~,
	\end{equation} where the first integral provides the mean number density of galaxies in the field, and 
	the second enhances this number according to the radial cross-correlation function, and integrates it 
	over the volume of the bin.

\item For every radial bin, we sample a Poisson distribution centered at the values $n_i$ to obtain an 
	integer number of galaxies. In most iterations, the total number of galaxies in the bins, and in the whole 
	halo, is zero since the average number of galaxies per bin  fluctuates within the range 
	$10^{-2}\gtrsim n_i \gtrsim 10^{-5}$, and the total number of  undetectable galaxies in the halo is 0.25.
	If the total number of galaxies is null, we repeat this step considering a new realization.
	
\item If the previous step results in one or more galaxies, we then assign them a luminosity (magnitude) 
	by using the inverse cumulative distribution function (ICDF) sampling method applied to the luminosity 
	function $\phi_{\rm H}(M)$.
	If the magnitude 
	assigned to the galaxy is smaller (the galaxy is brighter) than our detection threshold, we remove this 
	galaxy from the calculation and do not further consider it. If all galaxies are discarded because they 
	are detectable and maskable, we return to step 2.
	
\item	If there are undetectable satellite galaxies, we place them in the host halo. The radial distance is 
	set by the radial bin that the galaxies belong to, and we specify the positions on a sphere centered 
	at the quasar using two angles, $\Theta$ and $\Phi$, obtained by randomly drawing values from the 
	ranges $2\pi \gtrsim \Theta \gtrsim 0$ and $\pi/2 \gtrsim \Phi \gtrsim -\pi/2$. We then project the position 
	of the undetectable satellites on the plane perpendicular to the line-of-sight to obtain the impact 
	parameter of each source.
	
\item Finally, we transform the magnitude of every satellite to flux density and divide it by the area of 
	the radial annulus where the projected galaxies are placed, thus obtaining the surface brightness 
	values.
\end{enumerate}

 \begin{figure} 
\includegraphics[width=0.495\textwidth]{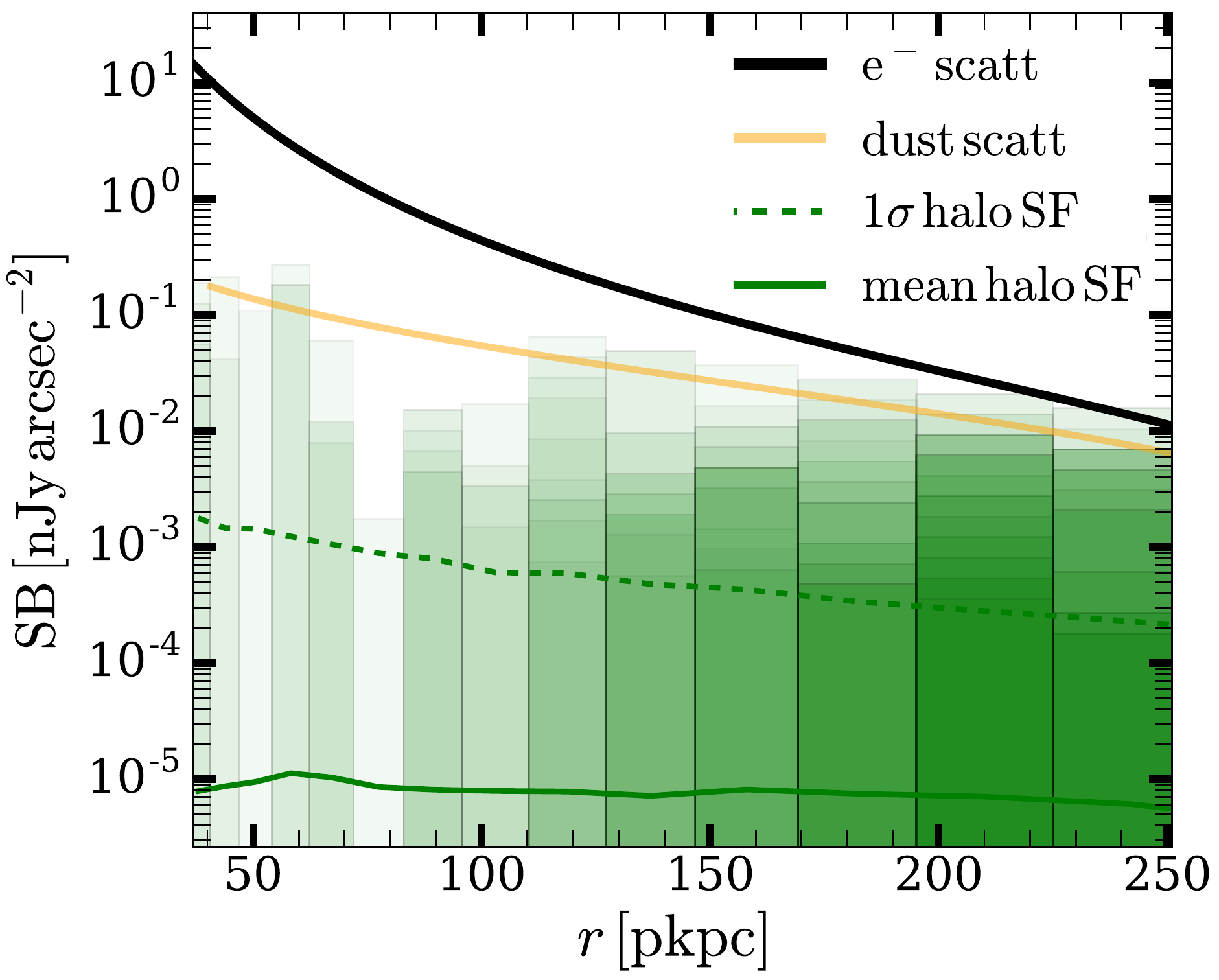}
\caption{ Surface brightness profiles from Figure \ref{fig:2dmap}, for 
electron ({\it black line}) and dust scattering ({\it orange line}). Each {\it green bar} denotes  
the surface brightness introduced by one undetectable satellite galaxy at its corresponding projected 
distance bin, typically representing one realization. The image shows around a 
thousand realizations overplotted, where in most cases the number of satellites is null. 
Most of the satellites reside within $\sim 150-250$ kpc and 
show surface brightness levels below the electron signal. The signal from satellites is $\gsim 1$ decade 
below that from electrons at projected distances $r\lesssim 100$ physical kpc. The {\it solid} and 
{\it dashed green lines} represent the mean and one-sigma deviation, respectively, of the surface 
brightness profile from satellites (halo star formation; SF) after $10^6$ realizations. A publicly available movie at 
\url{https://github.com/lluism/QSO_scattering} shows the iterative procedure for visualization.}
\label{fig:mcmc}
\end{figure}

   Figure \ref{fig:mcmc} illustrates the impact of the undetectable satellite sources on the electron- 
({\it black line}) and dust- ({\it orange line}) scattering surface brightness profiles from Figure 
\ref{fig:2dmap}. Every vertical {\it green bar} represents the surface brightness introduced by one 
undetectable satellite galaxy at its corresponding (projected) radial bin, after computing and overplotting one 
thousand realizations. In $\sim 75\%$ of the realizations there are no undetectable satellites and, when 
present, they mostly reside in the range within $\sim 150-250$ kpc from the center. At $r\lesssim 
100$ physical kpc, the electron-scattering surface brightness level 
is generally more than one order of magnitude higher than the 
brightest undetected sources, and the electron-scattering profile dominates the signal out to $\sim 200$ pkpc.
The {\it solid} and {\it dashed green} lines denote the mean and standard deviation values, respectively, 
of the surface brightness profiles for the undetected satellites after the full calculation with $10^6$ 
realizations. We stress that these profiles are much fainter than the typical surface brightness value 
introduced by the individual satellites because $\sim75\%$ of the realizations contribute to the calculation 
with zero satellites, i.e., null surface brightness, driven by the average number of $0.25$ galaxies per halo. 
Finally, the contamination will generally only affect one spurious impact parameter bin since this is the typical 
value of satellites per halo when they are present, which implies that satellite sources are not a strong 
contaminant to the overall electron-scattering profile. Given this result, we do not consider the effect of 
satellites in our further calculations.  A publicly available movie at \url{https://github.com/lluism/QSO_scattering} shows the 
iterative procedure for visualization.

\section{Observational Strategy}\label{sec:obs}

This section presents a detailed discussion of our observational approach, which aims to maximize
the signal-to-noise ratio of the electron scattered quasar radiation.

\begin{figure*} \center
\includegraphics[width=1\textwidth]{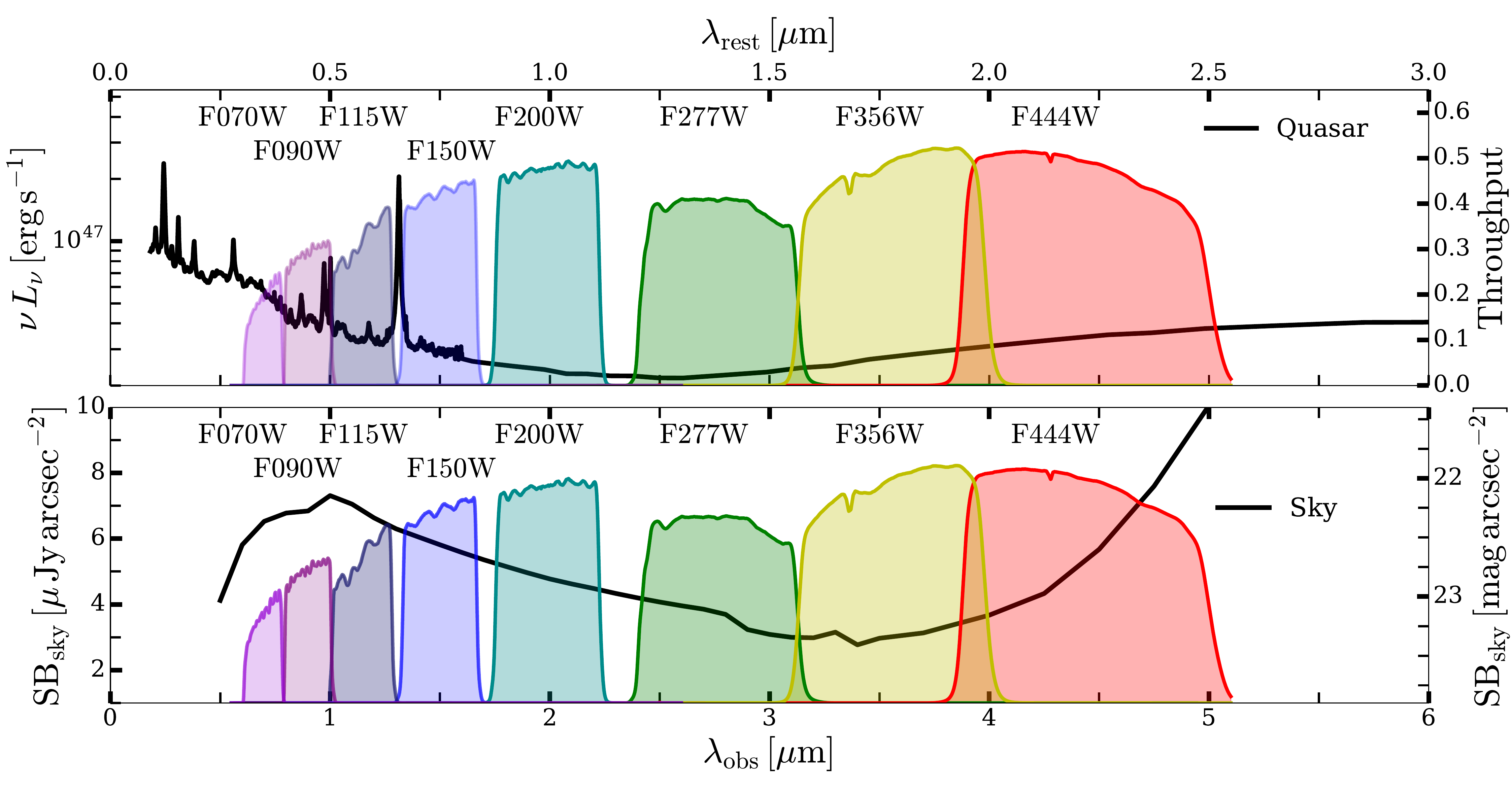}
\caption{{\it Upper panel:} Quasar SED at the observer and source ($z=1$) frames, with broad band NIRCam 
filters superposed on it. The left axis denotes the quasar luminosity and the right axis the total system 
throughput considering each filter. {\it Lower panel:} Surface brightness with wavelength for the 
background sky.}  
\label{fig:filters}
\end{figure*}

We consider the Near Infrared Camera (NIRCam) on board JWST. 
The large aperture of JWST (25 m$^2$) provides higher sensitivity to low surface brightness
emission than HST, and its well characterized PSF guarantees that regions contaminated by the
central  hyper-luminous quasar emission will be minimized. Our primary setup uses the broad-band filter F356W,
centered at $3.568$ $\mu$m with band width BW$\,=0.781$ $\mu$m. We choose this filter in particular, because
it is broad, and the signal-to-noise (S/N) ratio scales  as S/N $\propto$ BW$^{1/2}$ (Eq.~\ref{eq:snr} 
and \ref{eq:Nobs}),
and the central wavelength of F356W coincides with the region where
the sky background is the faintest. This is clear from the {\it lower
  panel} of Figure \ref{fig:filters}, which shows the NIRCam broad-band filters superposed on a plot of the
surface brightness of the JWST sky background versus wavelength from
\cite{Krick2012}\footnote{\url{https://jwst-docs.stsci.edu/display/JPP/JWST+Backgrounds}}.

To maximize the S/N ratio of the electron scattering halo we will
target a hyper-luminous quasar. For the choice of redshift, we choose
$z=1$ as our fiducial value, motivated by the following points: ({\it
  i}) at lower redshifts, the possible presence of extended stellar
halos around the host galaxy could be a significant contaminant of the
electron scattering signal. Indeed, we show that this could be an
issue for a hypothetical observation of 3C 273 at $z=0.16$, which we
discuss in detail in \S~\ref{sec:3c273}. We expect the amplitude of the 
extended stellar halo emission to be smaller at higher redshift, $z\gtrsim 1$
due to lower stellar masses and $(1 + z)^4$ surface brightness dimming. 
({\it ii}) The metallicity of the gas, directly related to the expected
amount of dust, is higher at lower redshifts, thus enhancing the
possible dust contamination.
({\it iii}) Given the well known strong luminosity evolution of quasars with increasing redshift,
there are already hyper-luminous quasars at $z\sim 1$ that are an order of magnitude brighter
than the brightest local quasars like 3C 273 ($z= 0.158$).
 ({\it iv}) Finally, at $z\sim 1$ our filter
covers the rest-frame NIR region of the quasar spectrum (see Figure~\ref{fig:filters}), whereas at $z\gsim 2$ it
shifts into the optical/UV, which increases the dust scattering optical depth relative to the electron
scattering optical depth, resulting in potential 
contamination from
dust scattering, as well as nebular radiation. In this case, the effect of
dust is not due to the variation of the amount of dust but to the wavelength 
dependence of its absorption
cross-section, which is much larger at optical/UV wavelengths than at
longer NIR wavelengths.
The contamination from nebular radiation is also most
important when observing in the UV range, because many bright hydrogen
and metal recombination lines (as well as continuum) are present. 
We analyse the redshift dependence of the electron scattering emission in  more detail in \S~\ref{sec:snrz}.

In order to maximize the signal-to-noise ratio of the electron scattering signal we wish to target hyper-luminous 
quasars, generally found at high redshift \citep[e.g.,][]{Boyle1988}, with their number density peaking at 
$z\sim2$ and decreasing rapidly at higher redshift \citep{Boyle2000,Wu2010}. 
\cite{Stern2015} analysed large samples of quasars and found that the brightest objects 
(in terms of apparent magnitude) generally inhabit the range $1\lesssim z \lesssim 1.5$. These objects 
have $i$-band apparent magnitudes of $\sim15.5$ mag, which we adopt as our 
intrinsic quasar brightness. We compute the spectral energy distribution (SED) of this hyper-luminous 
quasar following the procedure described in \cite{Arrigonibattaia2015}. Specifically, redward 
of the Lyman limit we
model the spectrum by splicing together the composite spectra by \cite{Lusso2015},
\cite{Vandenberk2001} and \cite{Richards2006}, and normalizing the amplitude to obtain the 
desired magnitude. The template by \cite{Richards2006} is the most relevant for our calculations since we 
focus on the rest-frame NIR range of the quasar spectra. 
For energies above 1 Rydberg, which we have used for our nebular calculations with {\sc Cloudy}, we 
make use of power laws: from 1 to 30 
Rydberg, we assume a power law $L_{\nu}=L_{\nu_{\rm LL}}(\nu/\nu_{\rm LL})^{\alpha_{\rm UV}}$, 
with  $\nu_{\rm LL}$ and $L_{\nu _{\rm LL}}$ denoting the frequency and luminosity at the Lyman 
limit, respectively, and ${\alpha_{\rm UV}}=-1.7$. Above 30 Rydberg and up to 2 keV we change 
the power-law index to -1.65, and to -1 for the X-ray band from 2 to 100 keV. Above this value, the 
hard X-ray slope is set to -2 \citep[see][for details on these calculations and references]{Arrigonibattaia2015}.
The {\it upper panel} in Figure \ref{fig:filters} displays the quasar SED in the rest and observer frames, 
with the NIRCam wide-band filters superposed on the spectrum. The vertical right axis denotes the 
total system throughput considering each of the filters.

 \begin{figure*}\center 
\includegraphics[width=0.85\textwidth]{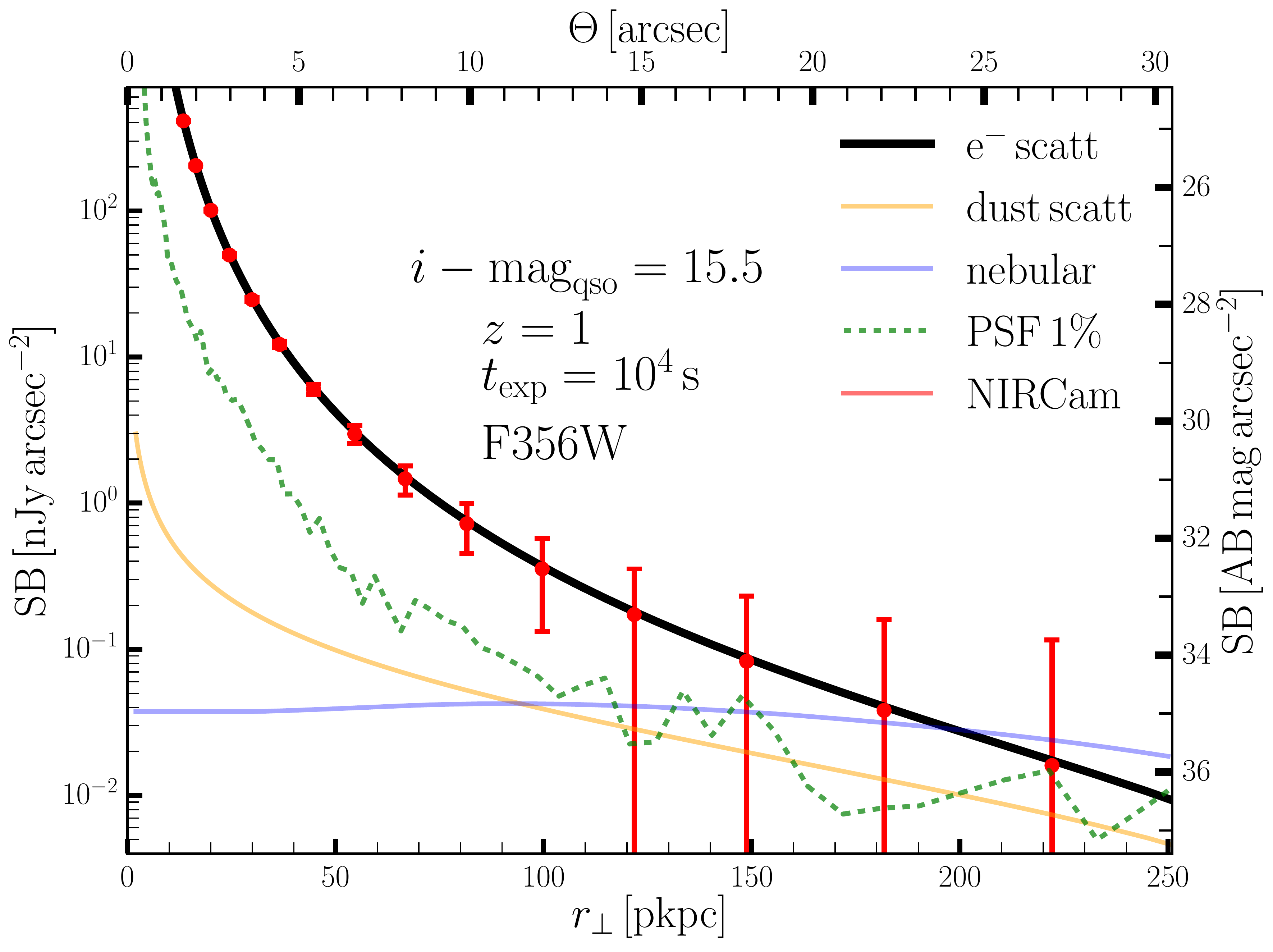}
\caption{Radial surface brightness profiles from the central quasar, for electron-scattered radiation 
({\it black line}), dust-scattered radiation ({\it orange line}), and nebular emission from 
the cool gas ({\it blue line}) around the host galaxy. The {\it dashed green} line denotes the 
residual profile of the PSF assuming that it is subtracted down to $1\%$ of its total value. 
The {\it red error bars} represent the observational uncertainty derived form the 
S/N ratio, calculated assuming an exposure time of $10^4$ seconds ($< 3$ h) with NIRCam onboard JWST, 
with the broad-band filter F356W, implying the observation of the quasar rest-frame NIR. This results 
indicate that the electron-scattered signal is detectable, and distinguishable from the other 
components, up to $\sim 100$ physical kpc from the center (beyond the PSF).} 
\label{fig:profile}
\end{figure*}

 Finally, using the aforementioned parameters, the detectability of the electron-scattering signal can be quantified.
 In all that follows we assume an exposure time of 10$^4$ s. The S/N ratio and the corresponding observational uncertainties for the electron-scattered surface brightness profile are computed according to
\begin{equation}\label{eq:snr}
{\rm S/N} = N_s / \sqrt{N_s+N_{\rm sky}+RN^2+N_{\rm PSF}}~. 
\end{equation}
Here, $N_s$ and $N_{\rm sky}$ are the azimuthally integrated photon number counts for the 
electron-scattered radiation and sky background, respectively, calculated from 
\begin{equation}\label{eq:Nobs}
N = \frac{f(\nu_{\rm obs})}{h_{\rm p}}\,\frac{\rm BW}{\lambda_{\rm obs}}\, { A_{\rm aper}}\, \eta\, t_{\rm exp} ~,
\end{equation}
where $h_{\rm p}$ is Planck's constant,  ${\lambda_{\rm obs}}=(1+z)\lambda_0$, is the observed frame
wavelength corresponding to the center of the filter, 
$A_{\rm aper}=25\,{\rm m}^2$
corresponds to the JWST aperture, and $\eta$ is the total system
throughput, which is shown on the upper right axis of Figure
\ref{fig:filters}. The term $f(\nu_{\rm obs}) = \int {\rm SB}(r_\perp,\nu_{\rm obs}) {\rm d}A_{r_\perp}$ is the 
source or sky flux density (in Jansky) computed by integrating the surface brightness profile within the 
area ${\rm d}A_{r_\perp}$ of logarithmically spaced radial annuli. The term $N_{\rm PSF}$ represents 
the photon count from the PSF, calculated using 
\texttt{WebbPSF}\footnote{\url{http://pythonhosted.org/webbpsf/}}, and convolving it with the total flux 
within the filter, assumed to be emitted by a point source. This term arises from considering 
the subtraction of the PSF in the analysis with a residual equal to $1\%$ of its total value. 
In practice, it represents the possible fluctuation between the 
actual PSF photon count and the average PSF determined in the subtraction. Because of the 
compactness of the JWST PSF compared to the large radii one would analyze for the electron scattering signal,
the noise contributed by the PSF term is small, and thus the details of PSF subtraction, and potential
systematics associated with it, are not a significant concern.

\section{Results}\label{sec:results}

The surface brightness profiles resulting from our calculations are presented in \S~\ref{sec:profiles}, 
and in \S~\ref{sec:snrz} we discuss the dependence of the detectability on redshift. An application of 
our formalism and the comparison to observations of the low-redshift quasar 3C 273 is performed in 
\S~\ref{sec:3c273}, and the profiles of the high-redshift quasar SDSS J152156+520238 are 
estimated in \S~\ref{sec:UM287}.

\subsection{Surface Brightness Profiles and Detectability}\label{sec:profiles}

Figure~\ref{fig:profile} illustrates our predicted surface brightness 
profiles and the expected S/N ratio for a hyper-luminous quasar at $z=1$.  
The electron-scattering 
profile is denoted by the {\it black line}, and is clearly above the 
dust ({\it orange line}) in the first hundred of physical kpc from 
the center of the host galaxy.  The vertical {\it red error bars} 
indicate the uncertainty in the NIRCam observations of the 
electron-scattered radiation, where the signal appears to be 
detectable (S/N $>1$) out to $\sim 100$ physical kpc (down to $\sim 
32.5$ mag arcsec$^{-2}$) from the center of the quasar host with an 
exposure of $10^4$ seconds ($2.78$ h). The {\it dashed green line} 
represents the profile of the filter PSF reduced down to $1\%$ of its original 
considering the PSF subtraction. 
Increasing the exposure time to 10 
hours would allow obtaining a detection above the noise out to $\sim 150$ pkpc. 
The {\it blue line} represents 
the nebular radiation resulting from the interaction between the 
radiation field of the quasar and the cool gas in the CGM as described 
in \S~\ref{sec:nebular}. 
We have tested that using the filters F444W (resulting in higher electron scattering 
signal but also a larger sky background) or F322W2 (extra-broad, covering the 
range with minimum sky background, i.e., $\sim 2.5 - 4\,\mu$m; Figure \ref{fig:filters})  
do not significantly alter  the results. 
This nebular component is not a significant 
contaminant of the electron profile below $\sim 150 - 200$ pkpc, and 
neither is the average profile of the satellite sources, whose level 
resides roughly two decades below the vertical scale of Figure~\ref{fig:profile}. Due to the small impact of 
satellites and nebular radiation, we henceforth consider only the electron and dust scattering profiles.

\subsection{Redshift Evolution of the Signal}\label{sec:snrz}

    The dependence of the electron-scattering emission S/N ratio on redshift is assessed in this section.  
In what follows, we focus on cosmological scaling, and will apply our formalism to two specific 
hyper-luminous quasars and discuss other effects, such as contamination from extended stellar 
halos and dust, in the next section. 

    First, we derive a simple analytical expression to gain insight into the evolution of the 
electron-scattering surface brightness and its detectability with redshift.  
We consider the surface brightness SB$(r_{\perp})$ resulting from the integration of the specific 
surface brightness from our previous calculations over a fixed observed-frame filter band-width 
$\Delta \nu_{\rm obs} = \nu_{\rm obs,2}- \nu_{\rm obs,1}$, where the two frequencies denote the 
filter limits, 
\begin{align}
  {\rm SB}(r_{\perp}) =& \int_{\nu_{\rm obs,1}}^{\nu_{\rm obs,2}}{\rm
      SB}(r_\perp,\nu_{\rm obs}) {\rm d}\nu_{\rm obs} \\ \nonumber 
      =   \frac{1}{(1+z)^3}& \tau_{\rm hot,0}\frac{1}{2\pi r_{\rm vir}^2}
      C(r_{\perp}\slash r_{\rm vir}) \int_{\nu_{\rm obs,1}}^{\nu_{\rm obs,2}}L_{\nu_0}{\rm d}\nu_{\rm obs} ~,
\end{align}
where $C(r_\perp\slash r_{\rm vir})$ is a geometric factor which
depends only on the radial profile slope of the hot gas $\alpha_h$. Assuming now a constant spectral
energy distribution, i.e., $L_{\nu_0}\nu_0={\rm const}$, and noting that
\begin{align}
  \int_{\nu_{\rm obs,1}}^{\nu_{\rm obs,2}}L_{\nu_0}&{\rm d}\nu_{\rm obs} =  \int_{\nu_{\rm obs,1}\slash (1 + z)}^{\nu_{\rm obs,2}\slash (1 + z)}\frac{L_{\nu_0}}{1 + z}{\rm d}\nu_0 \\ \nonumber 
 &= \frac{L_{\nu_0}\nu_0}{1 + z} \ln(\nu_{\rm obs,2}\slash \nu_{\rm obs,1}) \approx  \frac{L_{\nu_0}\nu_0}{1 + z}\frac{\Delta \nu_{\rm obs}}{\nu_{\rm obs}}~, 
\end{align}
we arrive at
\begin{equation}
  {\rm SB}(r_{\perp}) = \frac{C(r_{\perp}\slash r_{\rm vir}) }{(1+z)^4}
      \tau_{\rm hot,0}\frac{L_{\nu_0}\nu_0}{2\pi r_{\rm vir}^2}\frac{\Delta \nu_{\rm obs}}{\nu_{\rm obs}}~. 
\end{equation}

Thus, the surface brightness due to electron scattering
scales as the usual $(1 + z)^{-4}$ from cosmological surface
brightness dimming, times the quantity $\tau_{\rm hot,0}\slash r_{\rm vir}^2
\propto n_e\slash r_{\rm vir}$. Since the electron density $n_e
\propto (1 + z)^3$ and the virial radius $r_{\rm vir}\propto (1 + z)^{-1}$
(ignoring the other weak redshift dependencies in the equation for
$r_{\rm vir}$), we see that $\tau_{\rm hot,0}\slash r_{\rm vir}^2 \propto (1 +
z)^4$. Thus the electron-scattering SB$(r_{\perp})$ is redshift independent!  
The fact that the gas density increases with redshift as $(1 + z)^3$, and that the virial 
radius, which quantifies the size of the high-redshift  shock heated 
regions, decreases as $r_{\rm vir} \propto 1/(1 + z)$ cancels out the 
cosmological effect of surface brightness dimming. 

Next, we assess our ability to detect the electron-scattering signal with redshift. 
Assuming for simplicity that the detectability is background limited (i.e., $N_{\rm sky}$ dominates the 
noise in Eq.~\ref{eq:snr}), and integrating the surface brightness over both frequency and the angular
aperture enclosing the source, we obtain
\begin{equation}
\int {\rm SB}(r_{\perp},\nu_{\rm obs})\, {\rm d}\nu_{\rm obs} {\rm d}\Omega \,\approx\, {\rm SB}(r_{\perp})\Delta\Omega ~, 
\end{equation}
where $\Delta \Omega \simeq (r_{\rm vir}\slash D_A)^2$ and $D_A$ denotes the angular diameter distance.
Combining with Eq.~\ref{eq:snr}, 
\begin{align}
  {\rm S\slash N} \, \propto \, \frac{{\rm SB}(r_{\perp})}{\sqrt{{\rm
        SB}_{\rm sky}(r_{\perp})}} \sqrt{\Delta \Omega} \, \propto \,
  \frac{{\rm SB}(r_{\perp})}{\sqrt{{\rm SB}_{\rm
        sky}(r_{\perp})}}\left(\frac{r_{\rm vir}}{D_A}\right) ~.
\end{align}
Thus, for a fixed-luminosity (flat spectrum) source, the SB$(r_{\perp})$ is
redshift independent, and the S/N  ratio
scales as the angular size of the object, $\Delta \theta_{\rm vir} = r_{\rm vir} \slash D_A$. This
calculation was idealized in that we adopted a flat spectrum source, and imagined
integrating over a fixed observed frame frequency range. It thus ignores
the fact that the range of rest-frame frequencies that one probes
shifts blueward with increasing redshift. Nevertheless, it illustrates
the the main dependencies with redshift.

 \begin{figure*}\center 
\includegraphics[width=0.75\textwidth]{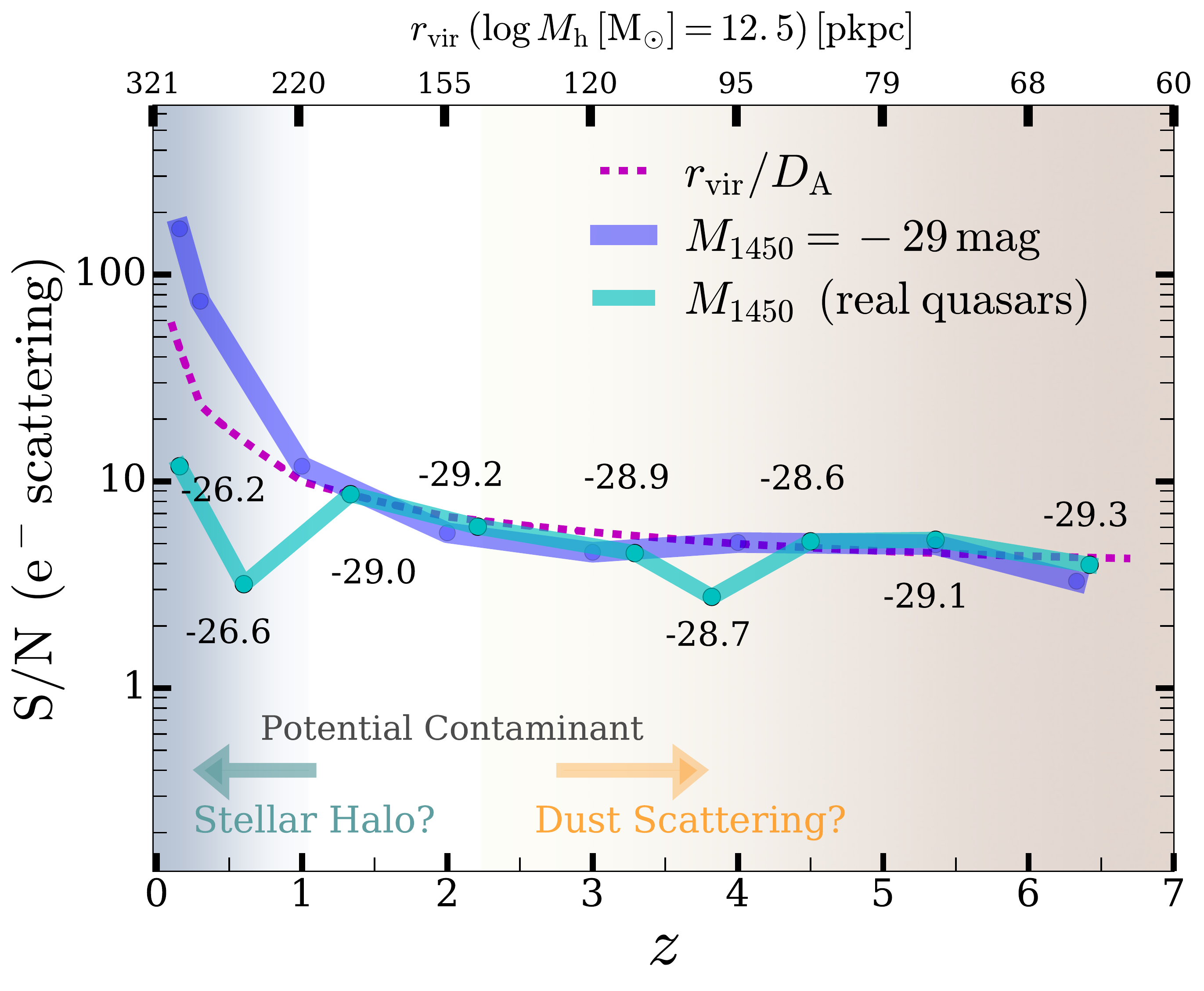}
\caption{Dependence of the detectability (expressed as the S/N ratio) on redshift for the electron-scattered 
emission, considering the default JWST NIRCam filter F356W, an exposure time of $10^4$ s, and the  
projected distance bin $0.3 \le r_{\perp}/r_{\rm vir} \le 1$. 
The {\it purple line} denotes the evolution for a fixed quasar brightness of 
$M_{1450}=-29$ mag, consistent with the absolute magnitudes of the brightest quasars in \cite{Stern2015} 
and \cite{Richards2006b} at $z > 1.3$, and the {\it dashed line} represents our analytical prediction as 
$r_{\rm vir}/D_A$, normalized to match the purple line at $z=4$ for comparison.
The {\it cyan line} shows the evolution for the brightest observed quasars, listed in Table \ref{ta:sch}, 
with their absolute magnitudes $M_{1450}$ shown in the plot. The detectability depends weakly on redshift, 
which allows measurements of the electron-scattered emission up to the Epoch of Reionization at $z\sim6.5$. 
The {\it blue gradient} illustrates the potential contaminant emission from extended stellar halos inhabiting 
the quasar host galaxies, important at redshifts $z \lesssim 1$, and increasing with color strength toward  
lower redshifts. The {\it orange gradient}  qualitatively  
represents the increasing impact of dust scattering on the electron-scattering signal with redshift, which 
starts being important in our models at $z\sim2.5-3$. We have assumed a constant dark matter halo mass 
of $\log\,M_{\rm h}[{\rm M_{\odot}}]=12.5$, and computed the corresponding virial radii at each redshift, 
shown in the upper horizontal axis.}
\label{fig:snrz}
\end{figure*}

We now more accurately calculate the values of the S/N ratio for the
electron-scattered emission at different redshifts and compare them
with our analytical results. We consider our default JWST filter
F356W, an exposure time of $10^4$ s, and the projected-distance bin
$0.3 \le r_{\perp}/r_{\rm vir} \le 1$ in order to avoid the area
contaminated by the central source PSF. We assume a constant dark
matter halo mass of $\log M_{\rm h}[{\rm M_{\odot}}]=12.5$, and
compute the corresponding virial radius at each redshift. The {\it purple line} in 
Figure \ref{fig:snrz} represents the S/N ratio 
evolution with redshift for quasars with a constant UV absolute
magnitude $M_{1450} \simeq -29$ mag, consistent with the brightest SDSS
quasars in \cite{Richards2006b}, and in \cite{Stern2015} at redshifts
$z > 1.3$. The {\it dashed line} illustrates the evolution of the ratio
$r_{\rm vir}/D_A$ expected from our analytical calculation, normalized
such that the two curves coincide at $z=4$ for comparison. The steep
rise of the signal at $z<1$ is driven by the behavior of the angular
diameter distance and enhanced by the $(1 + z)^{-1}$ redshift scaling
of the virial radius. At higher redshifts, $z \gtrsim 1.5$, the angular
diameter distance begins to mildly decrease with increasing redshift,
resulting in a flatter evolution.  The differences between the purple
and dashed lines arise from the fact that for each redshift our filter
observes different parts of the rest-frame quasar spectrum, and
the quasar SED rises toward bluer wavelengths (Figure \ref{fig:filters}). 
The {\it cyan line} in Figure \ref{fig:snrz} denotes the S/N ratio considering the  
brightest observed quasars, listed in Table \ref{ta:sch}, with their $M_{1450}$ 
absolute magnitudes indicated in the plot. The two solid curves follow well our 
predicted evolution (dashed line), enabling the measurement of the electron-scattered 
quasar emission up to the redshifts of cosmic reionization at $z\sim6.5$. Despite 
the low brightness ($M_{1450}=-28.6$ mag) of the $z=4.5$ quasar, the signal to noise 
appears as high as that of brighter objects because the strong H$\alpha$ emission 
line falls at the center of our filter in this case. 
The {\it blue gradient} in Figure \ref{fig:snrz} illustrates the potential contaminant 
emission arising from the extended stellar halos inhabiting the quasar host galaxies, 
which appears to be important at redshifts $z \lesssim 1$ (see next section). The 
{\it orange gradient} represents the impact of dust scattering on the 
electron-scattering signal with redshift, which starts being important 
in our models at $z\sim2.5-3$ (see \S~\ref{sec:UM287}).  The strength of the color 
in the gradients illustrate qualitatively the potential increase of the contaminants effects. 

\begin{table}\center
	\begin{center}
	\caption{Observed quasars}	\label{ta:sch}
	\begin{threeparttable}
		\begin{tabular}{cccc} 
		\hline
		 Quasar	   				&$z$		&$M_{1450}\,{\rm (mag)}$	&Ref.\,\tnote{a} \\ \hline
		3C 273					&$0.158$	&$-26.2$	&$1$	\\
		SDSS J210001.24-071136.3  	&$0.600$	&$-26.6$	&$2$	\\
		PG 1634+706				&$1.334$ &$-29.0$	&$2$	\\
		SDSS J152156.48+520238.5	&$2.208$	&$-29.2$	&$2$ \\
		SDSS J090033.50+421547.0	&$3.290$	&$-28.9$	&$2$	\\
		SDSS J163909.10+282447.1     &$3.819$	 &$-28.7$	&$2$	\\
		SDSS J134743.29+495621.3     &$4.510$	 &$-28.6$	&$2$	\\
		SW J030642.51+185315.8      	&$5.360$	&$-29.1$	&$3$	\\
		SW J010013.02+280225.8      	&$6.326$	&$-29.3$	&$4$	\\
		\hline
		\end{tabular}
		\begin{tablenotes}
			\item[a] Sources: (1) \cite{Soldi2008}; (2) \cite{Stern2015}; (3) \cite{Wang2016}; (4) \cite{Wu2015}.  
		\end{tablenotes}		
	\end{threeparttable}
	\end{center}
\end{table}

In the next section, we perform detailed calculations for real objects at low and high redshifts, and further
explore the impact of potential contaminants.

\subsection{Application to Real Quasars}\label{sec:real}

Here we apply our formalism to two real hyper-luminous quasars, the radio-loud 
source 3C 273 at $z=0.158$ in \S~\ref{sec:3c273}, and the quasar SDSS 
J152156+520238 at $z=2.208$ in \S~\ref{sec:UM287}, and compare the results.

 \begin{figure*}\center 
\includegraphics[width=0.498\textwidth]{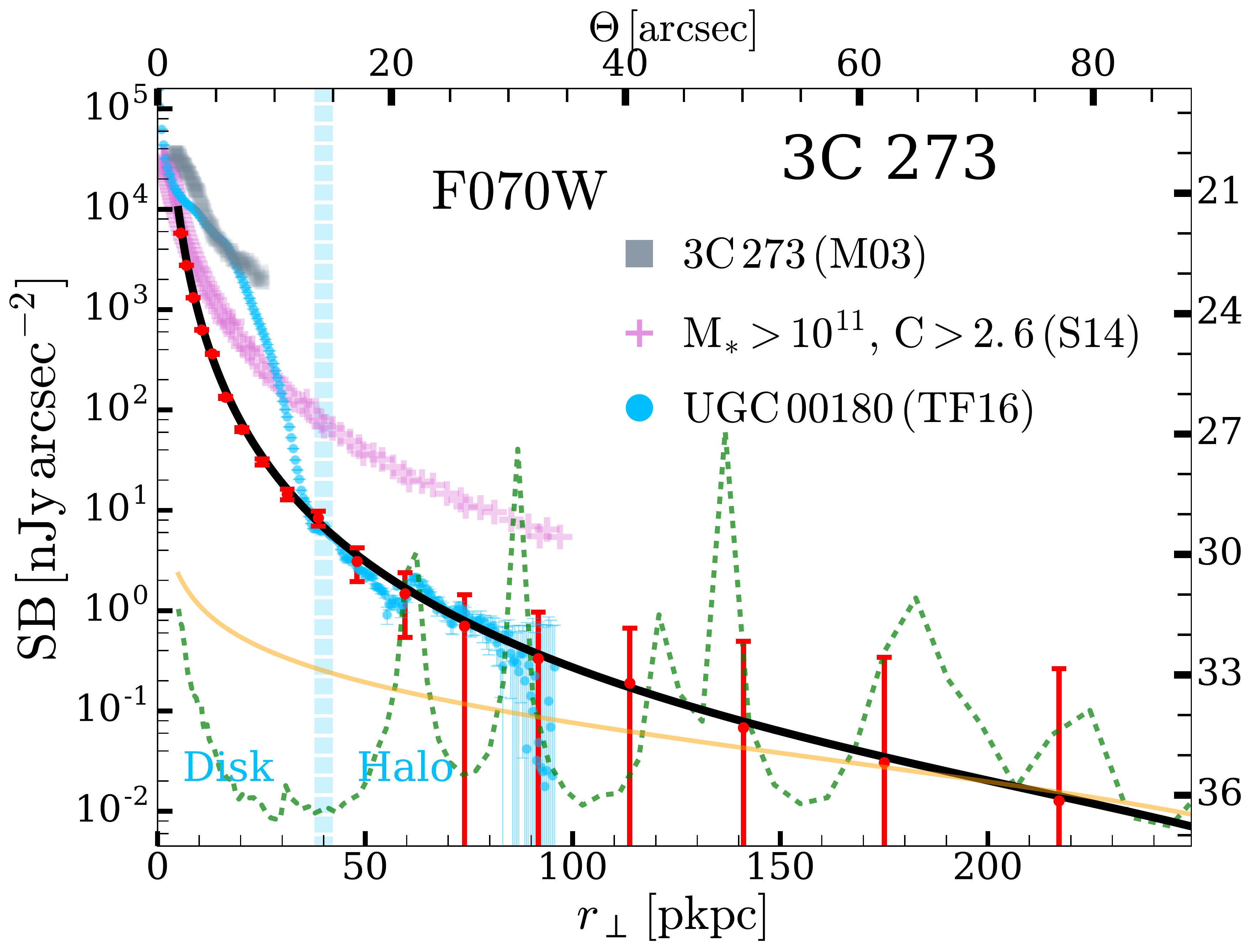}\includegraphics[width=0.495\textwidth]{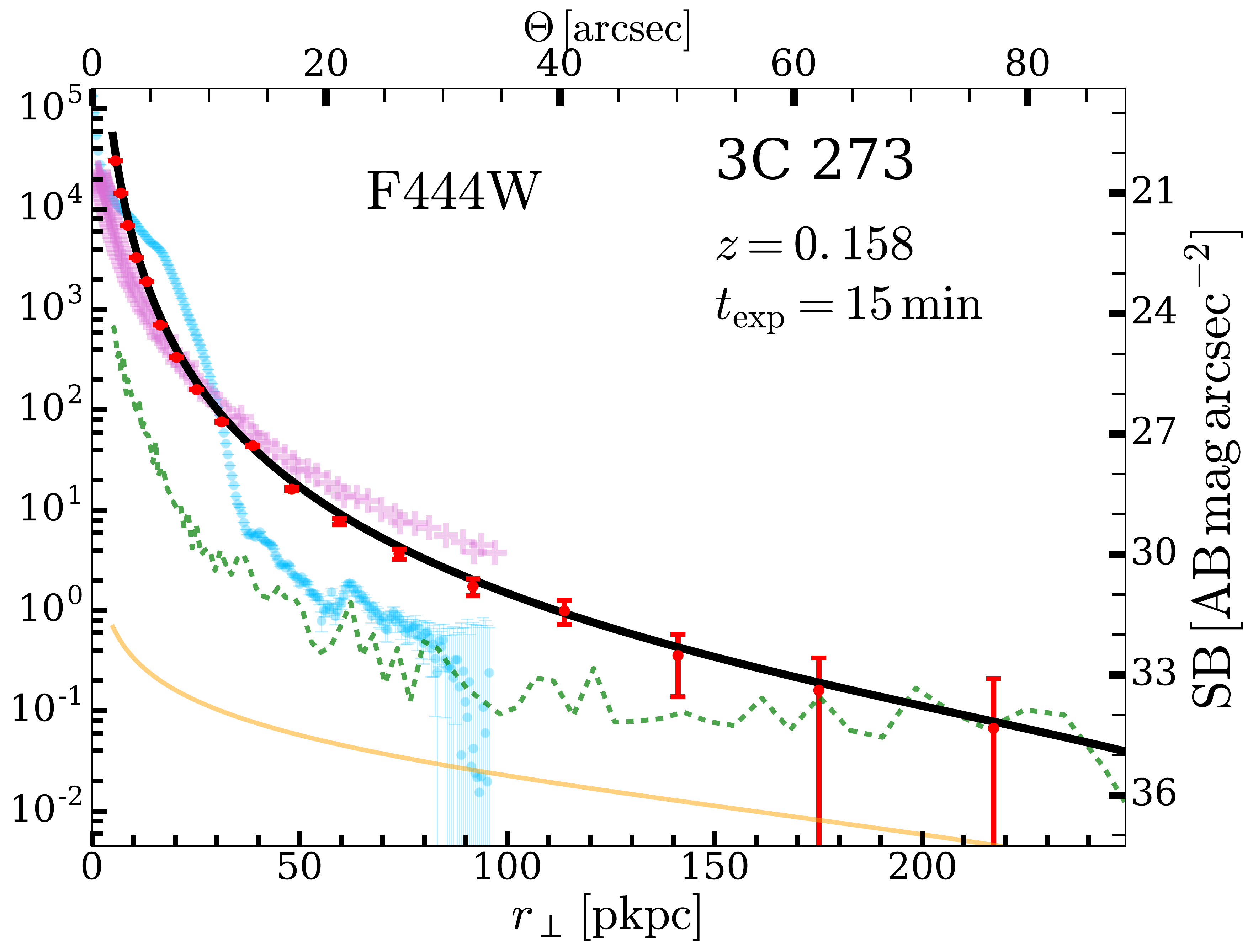}
\caption{Surface brightness profiles for the radio-loud galaxy 3C 273. {\it Left panel:} 
electron- ({\it black line}), dust-scattering ({\it orange line}), and $1\%$ PSF profiles as in previous figures but 
considering the NIRCam filter F070W, centered at $0.7\,\mu$m, 15 minutes of exposure time, 
and the 3C 273 continuum-only quasar spectrum template by \cite{Soldi2008}. The {\it gray squares} 
denote the HST-ACS  F814W ($I$-band) coronagraph data of 3C 273   by \cite{Martel2003}, and 
the {\it violet crosses} the $r$-band stack of SDSS galaxies in the ranges $0.06 \le z \le 0.10$, $M_\ast \sim 
10^{11} - 10^{11.4}\,{\rm M_{\odot}}$, and with concentration parameter $C>2.6$ by \cite{Souza2014}, 
consistent with isolated elliptical galaxies. We plot the D'Souza et al. data in the first 100 pkpc, where 
the uncertainties are small compared to the signal. For comparison, the {\it blue data} is the profile of the spiral 
Sab galaxy UGC 00180 observed in the $r$-band by \cite{Trujillo2016}, corrected by redshift dimming. 
The {\it vertical blue} line denotes the separation between the disk and the halo of UGC 00180 proposed 
by Trujillo \& Fliri. {\it Right panel:} same as in the left panel but with calculations performed at $4.4\,
\mu$m, with the filter F444W, and where we have corrected for the variation of the flux with wavelength 
(see text). In general, the stellar-halo profiles appear as a potential contaminant at this low redshift.}
\label{fig:3c273}
\end{figure*}

\subsubsection{3C 273 and Extended Stellar Halos}\label{sec:3c273}

     The radio-loud quasar 3C 273 \citep{Schmidt1963} is a well-studied nearby source, 
placed at $z=0.158$ ($749$ Mpc), with brightness $M_{\rm V} = -26.7$, and a 
bright extended radio jet
\citep[e.g.,][]{Uchiyama2006}.
The host is an elliptical E4 galaxy with 
(at least) four neighbouring galaxies within 150 pkpc  \citep[][see the 
review by \citealt{Courvoisier1998}]{Bahcall1996}.     

     Images of the inner parts of 3C 273 in most frequency bands appear saturated due to the 
brightness of the quasar. Therefore, studies of the host demand the use of coronagraphs 
to mask the central source \citep[e.g.,][]{Martel2003}, and accurate modeling and 
subtraction of the PSF \citep{Bahcall1995,Hutchings2004}, which requires the \emph{Hubble Space Telescope}'s
compact and stable PSF. \cite{Martel2003} used the HST Advanced Camera 
for Surveys (ACS) coronagraph to obtain surface 
brightness profiles out to $\sim 25$ kpc ($\sim 9$ arcsec)
in the $I$, $V$, and $g$ bands. Their observations 
suggested the presence of an extended stellar halo around the host galaxy, with no evidence of blue, 
young star-forming regions, and morphology similar to that of elliptical galaxies at large radii, 
but with a possible spiral structure and signatures of a merging event close to the center. 

Here we compare our calculations of the electron and dust scattering
surface brightness profiles with the observations of 3C 273 by Martel
et al. To get a handle on the expected extended spatial profile of
stellar emission, we also compare the average (stacked) surface
brightness profile of massive SDSS early-type galaxies by \cite{Souza2014}, and the deep
observations of a local spiral by \cite{Trujillo2016} for comparison.

Because the observational data cover the wavelength range $\sim 0.6 - 0.8\,\mu$m, we now 
calculate the scattering profiles considering the 
NIRCam filter F070W, centered at 0.7 $\mu$m, corresponding to a quasar rest-frame wavelength of 
$\sim 6000\angs$, and adopt an exposure time of 15 min.
For 3C 273's SED, we use the fitting profile from \cite{Soldi2008}, which extends from
ultraviolet to millimeter wavelengths (green line in their Figure 5).
We use this spectrum because it is specific for this object, but we have checked that using our default 
quasar template does not result in significant differences. 

The {\it black} and {\it orange lines} 
in the {\it left panel} of Figure \ref{fig:3c273} denote the computed electron- and dust-scattering emission profiles, 
respectively, and the {\it red error bars} the expected uncertainties in JWST measurements. The {\it gray squares} 
extending out to $\sim 9$ arcsec ($\sim 25$\,kpc) represent the HST/ACS F814W ($I$-band) 
observations of 3C 273 by \cite{Martel2003} 
without applying any correction, which are significantly above the expected scattering signals 
at impact parameters $\gtrsim 15$ kpc. 

     To obtain a more general comparison which extends out to larger radii, we also show 
the profile resulting from the stacking of massive SDSS early-type galaxies in the $r$-band from \cite{Souza2014}. We 
select their stack of galaxies inhabiting the redshift range $0.06 \le z \le 0.10$, the stellar mass range 
$M_\ast \sim 10^{11} - 10^{11.4}\,{\rm M_{\odot}}$, and concentration parameter $C>2.6$, which 
represents high-mass isolated central galaxies \citep{Wang2012}, with ellipticity consistent with 
that of the LRG galaxy sample in \citealt{Tal2011} (typical central ellipticals in galaxy groups).
This profile is represented by the {\it violet crosses} and appears brighter than that of the electron
scattering by a factor of around five at impact parameters $\gtrsim 50$ pkpc, suggesting that the extended stellar halo of
low-$z$  massive early-type galaxies, if present, can in fact dominate the extended emission. We plot the data by 
D'Souza et al. below the first 100 pkpc from the center, because the uncertainty in the profile 
beyond this point rapidly becomes very large, and possibly limited by systematics in their stacking procedure.

Because 3C 273  does not belong to a galaxy group, and given the evidence for spiral structure 
in the inner parts of its host galaxy \citep{Martel2003},
we also compare to the profile 
for the spiral (Sab) galaxy UGC 00180 observed in the $r$-band by \cite{Trujillo2016}, 
representing one of the deepest observations of extended emission around local galaxies.
The galaxy UGC 00180 is similar to the Andromeda galaxy (M31), with 
a stellar mass of $M_{\star}\sim1.3\times10^{11}\,{\rm M_{\odot}}$, and at $z=0.0369$. In this case, 
we correct the original Trujillo \& Fliri data for the
$(1 + z)^4$ surface brightness dimming effect due to the different redshifts 
of the sources, and we plot it as {\it blue points} and {\it error bars}. The UGC 00180 profile is 
higher than that of electrons for the first $\sim 30$ kpc and comparable at larger distances. This 
transition point corresponds to the separation between the disk and the halo of UGC 00180 
({\it dashed vertical line}) noted by Trujillo \& Fliri. In view of this comparison, we conclude that 
the signal from spiral galaxies can overwhelm that of electron scattering in the central regions 
of the galaxy. In the halo, even though the stellar halo of spirals is typically fainter than for 
ellipticals, it can still reach levels comparable to the electron scattering signal.

We perform a similar comparison, now considering the mid-IR part of the quasar spectrum.  
We wish to observe the reddest possible part of the galaxy spectrum, where the possible contamination from 
nebular radiation is expected to be smaller than in the previous calculation, and the rest-frame wavelength 
corresponds to the Rayleigh-Jeans tail of the stellar emission, thus reducing the impact of the extended stellar 
halo. For this purpose, we use the 
broad-band filter F444W, centered at $4.4\,\mu$m, corresponding to the quasar rest-frame wavelength 
of $\sim 3.8\,\mu$m, and the same values as before for the other parameters. These scattering profiles are 
shown in the {\it right panel} of Figure \ref{fig:3c273}. The {\it violet crosses} indicate the 
profile for elliptical galaxies from \cite{Souza2014}, but now correcting the D'Souza et al. $r$-band 
measurements to values appropriate to mid-IR observations with F444W. 
Specifically, we assume that the average SED of
D'Souza et al. galaxy sample is well represented by the spectrum of the elliptical E4 galaxy (same type as 3C 
273) NGC 0584 in \citealt{Brown2014} (their Figure 9), which covers the wavelength range of 
interest, $\sim 0.15 - 30\,\mu$m.
According to the SED by Brown et al., $\lambda_{r{\rm-band}}\,f_{r{\rm-band}}/\lambda_{4.4}\,f_{4.4}=10$, 
which we use to re-scale the D'Souza et al. measurements to $4.4\,\mu$m. To obtain the specific 
surface brightness, however, we have to also account for the change in wavelength so that 
${\rm SB}_{\nu,4.4}= \lambda_{r{\rm-band}}/\lambda_{4.4} \,{\rm SB}_{\nu,r{\rm-band}}$, altogether resulting  
in small variations of the stellar-halo surface brightness profile compared to that in the left panel ($\sim 26\%$).
This small difference between the surface brightness at the two wavelength ranges considered here 
is consistent with the findings by \cite{Temi2008}, who found similar surface brightness 
levels for the mid-infrared $J$, $H$, $K$, $3.6$, $4.5$, $5.8$, and $8.0\,\mu$m passbands in the 
stack of 18 local elliptical galaxies.
Interestingly, these authors also found that the surface brightness differences between bands remain 
almost constant with the distance from the center of the galaxy (their Figure 2).
The corrected surface brightness profile from D'Souza et al.  still
overlaps with that of electrons, indicating that emission from an
extended stellar halo could still dominate even at the reddest mid-IR
wavelengths.  The {\it blue points} and {\it error bars} show the data
for UGC 00180 from \cite{Trujillo2016}, again corrected for the
difference in redshift, and now also rescaled to $4.4\,\mu$m. For the
latter we use the SED of the Sa spiral galaxy NGC 5953 in Brown et al.,
which indicates that
$\lambda_{r{\rm-band}}\,f_{r{\rm-band}}/\lambda_{4.4}\,
f_{4.4}=8$. This analysis suggests that the electron scattering
emission should dominate over the stellar emission at distances
$\gtrsim 25 - 30$ pkpc from the center, if the galaxy hosting 3C 273 is a spiral
galaxy like UGC 00180.

In conclusion, for the brightest nearby quasar 3C 273, the presence of
a stellar halo appears to be a potential contaminant for the
electron-scattering signal, but large variations in the stellar
surface brightness profiles, as well as in the shape of the spectra
between galaxies of the same type can exist
\citep[e.g.,][]{Mannucci2001}, and detailed analyses, which are beyond the scope
of our current work, should be carried out to assess the detectability
of other low-$z$ quasars.  Note, however, that while 3C 273 has a large
apparent magnitude $i\simeq 13$ mag, its absolute magnitude is just
$M_{1450} = -26.2$ mag (using the \cite{Soldi2008} et al. SED fit), which
is a factor of $\sim 30$ fainter than the hyper-luminous quasars at $z
> 1$ which have $M_{1450}\simeq -29$ mag (see Figure~\ref{fig:snrz}).
Thus, given that extended stellar halos are comparable to the expected 
electron scattering SB at $z \sim 0.16$ around 3C 273, we expect the 
extended stellar emission to be much fainter than 
electron scattering at higher redshifts for two reasons. First, at higher-$z$ 
the (rest-frame) extended stellar halo SB will be at most comparable 
(and possibly lower) than what we have assumed for 3C 273, under the 
plausible assumption that this emission scales with $M_\ast$ of 
the host galaxy, given the high values of $M_\ast \sim 10^{11}\,{\rm 
M_{\odot}}$ that we considered in Figure~\ref{fig:3c273} and that stellar masses 
are lower at higher redshift. But the 
quasars at higher-$z$ are $\sim 30$ times brighter, boosting the electron
scattering signal by the same factor. Second, whereas the extended
stellar halo emission will redshift away due to the strong $(1 + z)^{-4}$
scaling of cosmological SB dimming, we showed in \S~\ref{sec:snrz} that
the electron scattering SB is redshift independent. In summary,
although extended stellar emission will likely complicate efforts to detect
electron scattering emission from 3C 273, we expect contamination from stars
 to be much less important around hyper-luminous quasars at $z > 1$. Lastly, we note
that even at $z < 1$ it should be straightforward to assess whether extended stellar halos are a significant
contaminant by simply obtaining images of fainter quasars for which the electron scattering signal is expected
to be undetectable. 

Finally, we stress that while Thomson scattering is wavelength
independent, the optical depth to scattering by dust particles for
shorter wavelength (UV) photons is much higher than at redder IR
wavelengths, due to the increase of the dust absorption cross-section
with decreasing wavelength \citep[e.g.,][]{Pei1992}. This effect is
visible in Figure \ref{fig:3c273}, where the dust SB in the {\it right
  panel} (F444W probing rest-frame 4$\mu$m) is a factor
between $\sim 4-5$ lower than in the {\it left panel} (F070W probing rest-frame $\sim 6000\angs$). 
Observing the rest-frame IR is thus beneficial to
minimize the undesired contamination by dust.

\subsubsection{Hyper-Luminous Quasars at $z \gtrsim 2$}\label{sec:UM287}

We now estimate the scattering surface brightness profiles for the hyper-luminous 
SDSS quasar J152156.48+520238.5 at redshift $z=2.208$ \citep{Schneider2005}. This object has  
an apparent $i$-band magnitude $m_i=15.323$ mag ($M_{1450}\simeq -29.3$ mag), being the brightest 
quasar in the SDSS catalog at this redshift \citep{Stern2015}. 

\begin{figure}\center 
\includegraphics[width=0.495\textwidth]{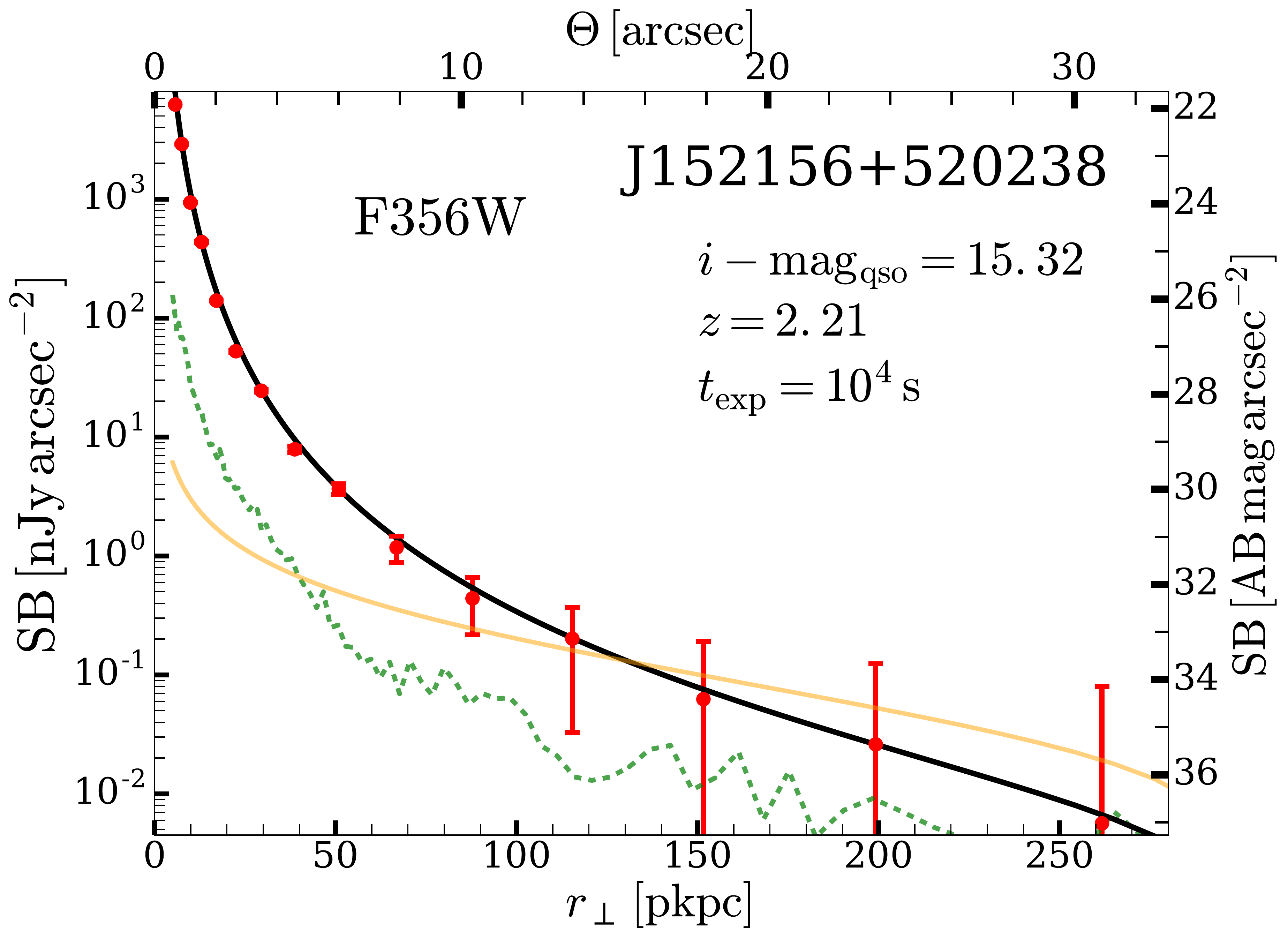}
\caption{Surface brightness profiles as in Figure \ref{fig:profile} but for the $z=2.208$  
quasar SDSS J152156+520238, with an exposure time of $\sim 3$ hours.}
\label{fig:sdssqso}
\end{figure}

Figure \ref{fig:sdssqso} displays the resulting profiles with our
default observational settings detailed in \S~\ref{sec:obs}, where the
electron scattering signal appears detectable out to $\sim 150$
pkpc. At distances beyond $\sim 115$ pkpc, however, the dust
scattering emission profile overwhelms that from electrons and the two
signals are indistinguishable beyond $\sim 90$ pkpc. The high dust
emission level results again from the wavelength dependence of the
dust scattering (absorption) cross-section, and the fact that at $z \simeq 2.2$
we probe bluer rest-frame wavelengths, $\sim 1.1\mu$m, than in our fiducial example
at $z\simeq 1$ (rest-frame $\sim 1.8\mu{\rm m}$, see Figure~\ref{fig:profile}). 

According to Figure~\ref{fig:snrz}, targeting comparably luminous
existing quasars at even higher redshifts ($z > 2$) would imply a
still higher S/N ratio for the electron-scattered emission. However in
practice, becasuse the dust optical depth increases toward bluer
rest-frame wavelengths, and hence toward higher redshifts in a fixed
observed frame filter, the separation between the electron and dust
scattering signals beyond $\sim 100$ pkpc could be challenging. Dust
emission dominates these profiles at distances beyond $\sim 80$ ($\sim
55$) pkpc at $z=3$ ($z=4$).  Note however, that our calculations
implicitly assume that the dust optical depth $\tau_{\rm cool, 0} \propto n_{\rm H, cool, 0}r_{\rm vir} \propto (1 + z)^4$,
the same scaling as scattering by electrons in the hot phase, becuase in our
cool gas model $n_{\rm H,cool,0}\propto (1 + z)^3$, just like the electron density
$n_{\rm e}$ in the hot phase. While arguments
based on gravitational collapse, virialization, and shock-heating
imply that the hot-phase density has to increase as $n_e \propto (1 + z)^3$, the
redshift scaling for the cool-phase gas density is much less clear, given that the
physical processes giving rise to the cool gas in the quasar CGM are
poorly understood \citep[e.g.,][but see \citealt{FaucherGiguere2015}]{FumagalliHennawi2013}. 
It is thus possible that the cool-phase density
does not track the redshift evolution of the mean density of the
Universe, which would imply significantly lower dust emission at high
redshifts. That said, by analyzing observations of multiple
filters covering a broad range of rest-frame wavelengths, one should be able to 
use the wavelength dependence of the signals to determine whether the scattering medium is
electrons versus dust, since the former results in a extended emission following the
SED of the quasar, whereas the latter follows the SED of the quasar multiplied by the
reddening law.


\section{Discussion}\label{sec:discussion}

We discuss the limitations and caveats of our proposed 
formalism in \S~\ref{sec:caveats}, and  alternative observational approaches in 
\S~\ref{sec:ground}.

\subsection{Caveats and Limitations}\label{sec:caveats}
    
     We explore the effects of considering different density gradients, such as that for the CGM of 
Milky Way type galaxies recently argued by \cite{Singh2018}, in \S~\ref{sec:singh}, discuss 
the redshift dependence of the electron density in \S~\ref{sec:reddis}, 
and the possible quasar obscuration and flickering in \S~\ref{sec:obscur}. 

\subsubsection{The Hot Gas Density Profile in the CGM}\label{sec:singh}

Using a Monte Carlo Markov Chain (MCMC) approach applied to the X-ray
and tSZ stacking results by \cite{Anderson2015} and \cite{Planck2013},
respectively, \cite{Singh2018} estimated the electron temperature and
gas fraction in the hot CGM phase of massive galaxies at $z~0.1-0.2$. 
Assuming a power law, their
calculations favor a radial dependence for the electron density of the
form $n_{\rm e}\propto r^{-1.2}$, which implies a power-law index
approximately a factor of two lower than our adopted value of
$\alpha_h = 5\slash 2$. This flatter density
profile would result in an overall improvement for the detection of the
electron-scattered profile at large radii in our fiducial calculations
at $z=1$. Although the signal decreases in the first few tens of pkpc
(by a factor of $3-4$ at $\sim 40$ pkpc), it still remains above that
of dust, and the S/N ratio is sufficiently high to allow accurate
measurements. Furthermore, the flattening enhances the electron
profile by a factor of $5-7$ at distances above $\sim 200$ pkpc,
extending the detected distances from $\sim 100$ out to $\sim 150$
pkpc. 

The differences between the profiles in Singh et al. and those in Nelson et al. 
might be attributed to the difference in redshift in the two studies, 
$z\sim0.1-0.2$ and $z=2$, respectively, and the different halo masses. Furthermore, 
one result arises from observations while the other is from simulations. 
The observation of the Thomson scattering emission and the 
subsequent modeling of the density profiles will be a useful tool to shed light on these 
dependencies, as well as for testing the prescriptions included in numerical simulations, 
specially those concerning complex feedback processes that can affect the properties of 
the hot halo phase.

\subsubsection{The Redshift Dependence of the Electron Density}\label{sec:reddis}

For the calculation of the redshift evolution of the electron scattering signal, we have 
considered that the electron density in the halo scales with redshift in the same way as the 
mean cosmic density, $(1+z)^3$. However, several models suggest that the halo evolution can be 
flatter, accounting for a more effective cooling at high redshifts due to 
the higher gas densities, that results in larger cool gas fractions \citep[e.g.,][]{Maller2004,
Sharma2012}. We do not explore further models for the electron density evolution, but it is 
important to keep these effects in mind for future comparisons with real observations.   

\subsubsection{Quasar Obscuration or Intermittent Emission}\label{sec:obscur}

Our calculations assume that the quasar radiation is emitted isotropically but, in reality, quasars are 
believed to be  surrounded by a thick dusty torus that will reduce the flux of UV/optical photons into the quasar CGM
by a factor $\Omega/4\pi$, where $\Omega$ is the solid angle that is unobscured by dust \citep[e.g.,][]{Antonucci1993,
Urry1995,Zakamska2005}. Obscured or Type II quasars represent the cases when the orientation of the object 
results in attenuation of the accretion disk and broad line region from our vantage point, and
in general, only radio, X-ray, and IR emission escape  the central regions toward the observer. In contrast, 
UV-bright Type I quasars are the cases where the accretion disk and broad line region are visible from our perspective.
The possible obscuration adds uncertainty to the measurement of the baryonic content from electron scattering
because it reduces the total flux of UV/optical photons into the CGM, reducing the expected surface brightness profiles. Thus
obscuration effects are  degenerate with determination of the electron density. 
However, the opening angle can be constrained because it is directly related to the fraction of
hyper-luminous  quasars that are obscured, i.e., $f_{\rm obscured}=1-\Omega/4\pi$. Observations of low-luminosity quasars 
suggest a fraction $f_{\rm obscured}=0.5$, implying an opening angle $\equiv \Omega = 2\pi$ \citep[e.g.,][and 
references therein]{Lusso2013}, but at high luminosities the opening angle has been debated. 
Some works argue that the so called `receding torus effect' implies that the most luminous sources  
are totally unobscured \citep[e.g.,][]{Honig2011}, while other studies based on number counts 
of obscured quasars claim that even the hyper-luminous quasars can suffer from obscuration, although 
these results are subject to large uncertainties \citep[e.g.,][]{Assef2015}. 

If quasars only  emit their UV/optical radiation into $2\pi$ steradians, the surface brightness profiles are 
reduced by a factor of two, not 
strongly impacting their detectability. This uncertainty affects, however, the interpretation of the 
observations due to the degeneracy with the electron density. Observations at long wavelengths, 
$\gtrsim 4\,\mu$m, reduce this concern because these wavelengths are much less sensitive to 
dust absorption and are essentially emitted isotropically. Furthermore, it may be possible to constrain the opening
angle of hyper-luminous quasars by studying the Lyman-$\alpha$ forest around them using background sightlines, exploiting
the so-called transverse proximity effect \citep[e.g.,][]{Schmidt2016,Schmidt2017}. 

Observations of the diffuse scattered emission around a luminous Type II (obscured) quasar would be 
interesting because the obscuring torus acts like a natural coronagraph and enables the study 
of the host halo and scattering signal at distances much closer to the central source. At the rest-frame
$\sim 1.8\,\mu$m wavelengths considered for our fiducial source at $z\sim 1$, 
the scattered emission in the central regions of the halo would be reduced by a factor of $\sim 1.5$ dex, which represents the flux
ratio between a Type I and II at this wavelength 
\citep[see figures 2 and 3 in][]{Honig2011}. At larger distances, the signal would be dominated by the (unobscured)
$\sim 1.5$ dex brighter radiation emitted in other directions and scattered into our line-of-sight,
although it may be difficult to quantify the Type I luminosity precisely, and therefore, the electron density.

Finally, our calculations assume that the quasars emit their radiation continuously over a timescale
sufficiently long that time-delay effects between radiation emitted at the same time but in different directions 
do not impact our results. The maximum time delay compared to photons emitted directly toward the observer 
will be for those photons emitted in a direction antiparallel to the line-of-sight toward the observer and then 
back-scattered by the electrons. For example, considering scattering off the free electrons at a radial 
distance of 100 pkpc from the source implies a maximum time delay of 
$6.5\times10^5$ yr, twice the crossing time.  If quasars flicker on timescales shorter
than this \citep[see, e.g.,][]{Eilers2017}, time-delay effects must be considered.
For the case of dust scattering, the impact of time delay is less significant because most 
of the observed radiation arises from emission at small angles from the line of sight.
For simplicity, we have not taken into account such effects in our current calculations.
   
\subsection{Ground Based Observations}\label{sec:ground}

One of the benefits of considering observations with JWST is its compact and stable PSF. In ground-based 
observations, the subtraction of the PSF is a vital (not simple to achieve) requirement to obtain a detection 
\citep{Dejong2008,Trujillo2013,Sandin2014,Sandin2015}. For instance, \cite{Hutchings2004} made use of a 
coronagraph, together with HST data, to support and reinforce their ground-based observations of the host 
galaxy around the bright 3C 273 quasar.  However, ground-based telescopes have larger apertures
and are less oversubscribed than space-based facilities, so it is worth discussing observations
from the ground. Indeed, if an obscured hyper-luminous quasar were identified with intrinsic (unobscured) luminosity
comparable to the brightest Type-Is, then the obscuring torus acts like a natural coronograph, and it would
be highly interesting to pursue deep observations from the ground. 

The level of our predicted surface brightness profiles is challenging
but achievable with current 8m-class telescopes.  Although the
emission profile in Figure~\ref{fig:profile} is as bright as $\sim 29$
mag arcsec$^{-2}$ in the inner regions, it approahces $\sim 32.5$ mag
arcsec$^{-2}$ at 100 pkpc.  \cite{Trujillo2016} explored the limits of
low SB observations on 8m-class, and reached a SB limit of $\sim 31.5$
mag arcsec$^{-2}$ ($3\sigma$ in a $10\times 10$ arcsec box) in an
8-hour $r$-band integration. By azimuthally averaging (over annular
bins with radii $\sim 100\arcsec$ comparble to the scales we consider
here), they were able to probe down to surface brightness levels of
$\sim 33$ mag arcsec$^{-2}$, comparable to our signal at 100 pkc.  Also
recently, \cite{Buitrago2017} analysed HUDF data to study elliptical
galaxies and their careful treatment allowed them to reach surface
brightness levels of $\sim 31$ mag arcsec$^{-2}$. Future optical
instruments, such as those planned for the
TMT\footnote{https://www.tmt.org},
GMT\footnote{\url{http://www.gmto.org/}} and
E-ELT\footnote{\url{https://www.eso.org/public/teles-instr/elt/}}
telescopes with apertures of $\sim$ 30 meters, will probe to deeper
levels, although at IR wavelengths, observations from the ground will
not be competitive with the extremely low SB that can be achieved with
JWST, and its possible space-based successors such as 
LUVOIR\footnote{\url{https://asd.gsfc.nasa.gov/luvoir/}}.

Finally, the Dragonfly Telephoto Array\footnote{\url{http://www.astro.yale.edu/dragonfly/index.html}} 
\citep{Abraham2014} is a novel small ground-based instrument designed to reduce considerably the undesired 
scattered light in the telescope 
compared to usual reflective telescopes, which makes it capable of reaching surface brightness levels below 
$\mu_B=30$ mag arcsec$^{-2}$ with observations of $\sim10$ hours. Dragonfly is well suited for targeting 
diffuse and extended structures and it has already demonstrated its potential for these type of 
observations \citep[e.g.,][]{Vandokkum2015b,Vandokkum2015,Merritt2016}.

\section{Conclusion}\label{sec:conclusions}

   We have demonstrated the feasibility of observing diffuse electron-scattered radiation from a hyper-luminous 
quasar in the circumgalactic medium (CGM) of the host galaxy with JWST, which can be used to probe the 
physical properties of the warm and hot gas, and to quantify the baryonic content in 
these CGM phases. We have parameterized the central quasar, and the radiation sources and gas in the host 
halo following observational and numerical results. We have calculated the electron- and dust-scattered surface brightness 
profiles considering the respective scattering redistribution functions, and accounted for the 
radiation from satellites sources, nebular (recombination) radiation, and potential extended stellar halos.
Our findings can be summarized as follows:

\begin{itemize}
\item The surface brightness profile of the NIR radiation from a luminous quasar at $z=1$, scattered by the 
	free electrons in the warm and hot CGM of the host galaxy, is detectable up to $\sim 100$ kpc physical from 
	the central quasar (at a surface brightness level of $\sim 32.5$ AB mag arcsec$^{-2}$) with less than 3 
	hours of imaging 
	observations with NIRCam onboard the James Webb Space Telescope (JWST). This signal appears above 
	those of dust, recombination, and halo star formation (after masking the brightest satellite sources), and should
        also be at least a factor of ten higher than extended stellar halo emission.

      \item A positive detection of this electron-scattering signal would provide a direct measurement of the radial profile
        of the number density of free electrons, and therefore, the amount of baryons in the warm and 
	hot CGM phases in high-redshift halos. 

      \item The electron-scattering surface brightness is redshift
        independent, because warm/hot gas is denser at higher redshifts and because the halos
        are more compact. The detectability of the signal scales as the
        angular size of the virial radius, which is a very weak
        function of redshift for $z \gtrsim 1$.  This implies that the
        signal is detectable around hyper-luminous quasars out to
        above $100$ physical kpc from the central source up to the redshifts of the Cosmic 
	Reionization at $z\sim6.5$ with 10$^4$ s ($<3$ h) of observation.

      \item At $z\lesssim1$ where quasars are intrinsically much
        less luminous, the extended stellar halos, which have been
        detected around massive nearby galaxies, could dominate
        over the electron scattering signal. For the
        hyper-luminous quasars at $z \gtrsim 1$, however, this signal will
        be a factor of $\sim 30$ lower than the electron
        scattering emission.

      \item At $z\gtrsim 2.5$ the electron-scattering
        signal might be contaminated by dust scattering if the
        density of cool gas in the quasar CGM scales as $\propto (1
        +z)^3$, similar to expectations for the warm/hot phase. This
        potential increase in the dust contamination at higher redshift
        occurs because one probes bluer rest-frame
        wavelengths, which increases the dust scattering optical
        depth relative to that from electron scattering. However, it
        may be possible to determine the nature of the scattering
        medium by analyzing the color of the signal.

\end{itemize}

Our proposed method of using observations of extended
Thomson-scattered radiation from hyper-luminous quasars aims to open a
new and unique window for subsequent detailed studies of baryons in
galactic halos, independent of their temperature, probing the spatial
distribution of the predicted warm and hot phases which have been extremely difficult to
observe. Furthermore, the presence of hyper-luminous quasars out to
redshifts as large as $z\sim 6.3$ \citep{Wu2015}, coupled with the
redshift inedependence of the electron scattering surface brightness (and the weak
redshift dependence of the S/N ratio),
suggest that we may be able to probe halo baryons via electron scattering over
10 billion years of cosmic history, provided that scattering by dust is not
a major contaminant.
This approach does not suffer from the difficulties and limitations of
other techniques that make use of X-rays or the thermal
Sunyaev-Zel'dovich effect, and therefore, can be crucial for setting
constraints on the impact of quasar feedback, as well as for
quantifying the `missing' baryons inhabiting the CGM of massive halos up to the redshifts of cosmic
reionization.

While the present work demonstrates that electron scattering halos should be easily 
detectable in high-resolution sensitive
JWST images, obtaining a spectrum of the diffuse emission would provide important additional 
information. Because Thomson scattered
photons inherit a Doppler shift determined by the electron velocities, which are moving 
$\sqrt{m_e\slash m_p}\sim 40$ faster than the
virial velocity, an electron-scattered quasar emission line will be broadened by $\sim 
10^4\,{\rm km/s}$, which exceeds the intrinsic line-widths ($\sim 3000$ km ${\rm 
s^{-1}}$; \citealt{Loeb1998}). If this broadening is detectable via a spectrum
of the scattered line emission, it opens up the exciting possibility of directly measuring the 
temperature of the baryons in high-redshift halos. Furthermore, these imaging and spectroscopic 
observations could be complemented with polarimetry, which would
definitively prove that scattering is the source of emission, because of the high polarization 
resulting from scattering off of dust and electrons \citep[e.g.,][and references 
therein]{Zakamska2005}. We will address these questions in detail in a future paper.

\section*{acknowledgements}
We thank the referee for important comments on the distribution of gas inside halos and its 
dependence on redshift, that have improved the discussion of our results. 
The initial inspiration for this work grew out of a stimulating
discussion with Ski Antonucci, and we are grateful to him for many
valuable and detailed comments on quasars and scattering.  We thank
Brice M\'enard and Yi-Kuan Chiang for their critical comments on
extended stellar halos around massive galaxies, and Ignacio Trujillo,
Feige Wang, Fabrizio Arrigoni Battaia and Jonathan Stern for sharing
their data/scripts with us.  We also thank Gordon T. Richards and
J. Stern for helpful discussions about dust in quasar environments,
Mark Dijkstra, Avi Loeb, and Sebastiano Cantalupo for noting important
radiative processes, and Jordi Miralda-Escud\'e and Andreu
Arinyo-i-Prats for highlighting the impact of potential time-delay
effects.  The authors are grateful to the scientists at the MPIA in Heidelberg
and the members of the UCSB/MPIA ENIGMA group, as well as those at the CCAPP and Astronomy
department of the Ohio State University, for their kind hospitality
and enriching discussions.

\bibliographystyle{apj}
\bibliography{qso_halo}\label{References}

\end{document}